\documentclass[pra,reprint,preprintnumbers,
amsmath,amssymb,
aps, showkeys
]{revtex4-2}

\usepackage{graphicx}% Include figure files
\usepackage{dcolumn}% Align table columns on decimal point
\usepackage{bm}% bold math
\usepackage{amssymb}
\usepackage{amsmath}
\usepackage{color}
\usepackage{soul}
\usepackage{subfig}
\usepackage{appendix}
\usepackage{booktabs}
\usepackage{comment}
\usepackage[hyperindex,breaklinks]{hyperref}

\newcommand{\id}{1\!\!1}

\begin{document}
	
	\title{$\mathcal{PT}$-symmetric two-photon quantum Rabi models}
	
	\author{Yi-Cheng Wang$^1$}
	
	\author{Jiong Li$^{1,}$}
	\email{jiongli@zju.edu.cn}
	
	\author{Qing-Hu Chen$^{1,2,}$}
	\email{qhchen@zju.edu.cn}
	
	\affiliation{$^1$Zhejiang Key Laboratory of Micro-Nano Quantum Chips and Quantum Control, School of Physics, Zhejiang University, Hangzhou 310027, China \\
		$^2$Collaborative Innovation Center of Advanced Microstructures, Nanjing University, Nanjing 210093, China}
	
	\date{\today}
	
	\begin{abstract}
		We investigate two non-Hermitian two-photon quantum Rabi models (tpQRM) that exhibit $\mathcal{PT}$ symmetry: the biased tpQRM (btpQRM), in which the qubit bias is purely imaginary, and the dissipative tpQRM (dtpQRM), where the two-photon coupling is made imaginary to introduce dissipation. For both models, we derive exact solutions by employing Bogoliubov transformations. In the btpQRM, we identify spectral collapse at a critical coupling strength, with accompanying $\mathcal{PT}$ symmetry breaking that correlates with exceptional points (EPs) arising from coalescing eigenstates. We establish a direct correspondence between $\mathcal{PT}$-broken regions and the doubly degenerate points of the Hermitian tpQRM, and analyze the effects of qubit bias via an adiabatic approximation. In the dtpQRM, although no spectral collapse occurs, both EPs and Juddian-type degeneracies are present, with well-separated behaviors distinguished by parity conservation. Through biorthogonal fidelity susceptibility and c-product, we successfully identify and classify the nature of these two types of level crossings. Finally, we compare the dynamical evolution of both models, revealing fundamentally different pathways to steady states governed by their respective non-Hermitian spectral structures. Our results provide exact characterizations of $\mathcal{PT}$-symmetric non-Hermitian tpQRMs and may offer theoretical insights for future experimental realizations.
	\end{abstract}
	
	\maketitle
	
	\section{Introduction} 
	
	Non-Hermitian systems, which exchange energy with their environment, have recently attracted considerable attention due to the complex eigenvalue spectra typically  produced by their Hamiltonians~\cite{bender_making_2007, moiseyev_non-hermitian_2011, ashida_non-hermitian_2020}. Such Hamiltonians are employed across diverse domains, including cold atomic systems~\cite{dum_monte_1992, otterbach_dissipative_2014, lee_heralded_2014, li_observation_2019, li_non-hermitian_2022}, superconducting vortex systems~\cite{hatano_localization_1996, refael_transverse_2006, longhi_non-hermitian_2013}, and surface hopping~\cite{neria_quantum_1991, granucci_including_2024, jaeger_decoherence-induced_2024}. Several theoretical frameworks have been developed to explore these unconventional phenomena, including quantum trajectories~\cite{moiseyev_non-hermitian_2011, ashida_non-hermitian_2020, HM_Wiseman_1996}, non-unitary conformal field theory~\cite{li_observation_2019, mostafazadeh_pseudo-hermiticity_2002, mostafazadeh_pseudo-hermiticity_2002-1, heyl_dynamical_2013, fring_exact_2017}, and biorthogonal quantum mechanics~\cite{curtright_biorthogonal_2007, brody_biorthogonal_2014, tzeng_hunting_2021}. Remarkably, under certain conditions, a non-Hermitian Hamiltonian exhibiting parity-time ($\mathcal{PT}$) symmetry can possess a purely real eigenvalue spectrum~\cite{bender_real_1998, klaiman_visualization_2008, ruter_observation_2010, bender_pt_2015, el-ganainy_non-hermitian_2018, wang_observation_2021}. Exceptional points (EPs), where eigenvalues coalesce and become complex, signal the transition between $\mathcal{PT}$-symmetric and $\mathcal{PT}$-broken phases. Numerous studies have highlighted the unique effects of EPs in non-Hermitian physics~\cite{T_Stehmann_2004, liertzer_pump-induced_2012, doppler_dynamically_2016, li_exceptional_2023}.
	
	The quantum Rabi model (QRM) describes a two-level system (qubit) coupled to a single electromagnetic mode (oscillator) via dipole interaction, representing the most fundamental form of light-matter coupling~\cite{rabi_process_1936,braak_integrability_2011}. It can be implemented across a range of physical platforms, including cavity quantum electrodynamics (QED)~\cite{scully_zubairy_1997, braak_semi-classical_2016, meystre_quantum_2021}, circuit QED~\cite{niemczyk_circuit_2010, forn-diaz_ultrastrong_2017, forn-diaz_ultrastrong_2019}, and trapped ion systems~\cite{leibfried_quantum_2003, pedernales_quantum_2015, cai_observation_2021}. To account for phenomena observed in recent experiments and quantum simulations involving multiphoton processes, nonlinear generalizations of the atom-cavity interaction have been introduced. Among these, the two-photon quantum Rabi model (tpQRM) serves as a crucial platform for investigating novel effects in nonlinear quantum optics. As the coupling strength increases, distinct features emerge, offering promising applications beyond conventional dipole interactions, such as alternative schemes for quantum information processing~\cite{bertet_generating_2002, stufler_two-photon_2006, del_valle_two-photon_2010, cong_polaron_2019}. A striking feature of the tpQRM in the strong-coupling regime is the collapse of the discrete energy spectrum at a critical coupling strength. Beyond this point, the spectrum becomes continuous, and the corresponding wavefunctions are no longer normalizable~\cite{felicetti_spectral_2015, duan_two-photon_2016, ying_symmetry-breaking_2021}.
	
	In recent years, $\mathcal{PT}$-symmetric non-Hermitian semiclassical and quantum Rabi models have attracted increasing attention~\cite{Ben-Aryeh_2004, joglekar_pt_2014, lee_pt_2015, xie_exceptional_2018, lu_pt_2023}. \textcite{joglekar_pt_2014} and \textcite{lee_pt_2015} introduced a purely imaginary coupling constant into the semiclassical Rabi model, revealing gain–loss dynamics in two-level systems. Furthermore, the non-Hermitian Dicke model—which describes two-level cold atoms in an optical cavity with dissipative atom–field coupling—has also been explored~\cite{luo_quantum_2024}. Most prior studies of non-Hermitian QRMs have focused on linear cavity-photon coupling, including a purely imaginary bias~\textcite{lu_pt_2023} and a purely imaginary qubit-cavity coupling~\textcite{li_ptmathcal_2025}.	In this work, we investigate two variants of the $\mathcal{PT}$-symmetric non-Hermitian tpQRM, focusing specifically on a purely imaginary bias and dissipative coupling to the cavity mode.	
	
	The tpQRM has been exactly solved by \textcite{chen_exact_2012} using the Bogoliubov operator approach (BOA), revealing key features such as spectral collapse~\cite{felicetti_spectral_2015, duan_two-photon_2016, xie_quantum_2017, li_two-photon_2020, xie_double_2021, duan_unified_2022}. More recently, \textcite{braak_spectral_2023} reproduced Chen’s solution in Bargmann space, showing that it exhibits an explicit pole structure that determines the collapse point~\cite{chen_exact_2012}. In this work, we derive the exact solutions of the non-Hermitian tpQRMs using the BOA. Based on these solutions, we examine whether the characteristic spectral collapse—an essential feature of the Hermitian tpQRM—persists in its non-Hermitian extensions. Moreover, we analyze both the original doubly degenerate exceptional points and those emerging from $\mathcal{PT}$-symmetry breaking.
	
	The paper is organized as follows. In Sec. II, we briefly introduce the two $\mathcal{PT}$-symmetric non-Hermitian tpQRMs.	In Sec. III, using the Bogoliubov transformation, we derive the exact solutions for both models.	In Sec. IV, we analytically obtain the EPs for both models and the doubly degenerate points in the spectrum of the dtpQRM.	These two types of special points can be distinguished using biorthogonal fidelity susceptibility and the c-product. Qualitatively distinct dynamical behaviors in both models are discussed in Sec. V. Finally, conclusions are drawn in Sec. VI.

	\section{Two Non-Hermitian tpQRMs}
	
	The general tpQRM can be described by the Hamiltonian
	\begin{equation}
		H_{\mathrm{gtp}} = \frac{\epsilon}{2} \sigma_z - \frac{\Delta}{2} \sigma_x + \omega a^{\dagger} a + g \left[ a^2 + (a^{\dagger})^2 \right] \sigma_z, \label{H_gtp}
	\end{equation}
	where $a$ and $a^{\dagger}$ are the annihilation and creation operators of a single cavity mode with frequency $\omega$, $\Delta$ denotes the qubit energy splitting, $g$ is the two-photon qubit–cavity coupling strength, $\epsilon$ is the qubit bias energy, and $\sigma_{x,y,z}$ are the Pauli matrices. For convenience, we set $\omega = 1$ throughout this paper.
	
	In this work, we extend the tpQRM to two non-Hermitian, $\mathcal{PT}$-symmetric variants. In the first, referred to as the biased tpQRM (btpQRM), the qubit bias energy $\epsilon$ is replaced by a purely imaginary value $i\epsilon$. In the second, termed the dissipative tpQRM (dtpQRM), the coupling strength $g$ is replaced by $ig$, introducing a dissipative interaction between the qubit and the cavity field.
	
	\textsl{Biased Two-Photon Quantum Rabi Model:} The Hamiltonian of the btpQRM is given by
	\begin{equation}
		H_{\mathrm{btp}} = i\frac{\epsilon}{2} \sigma_z - \frac{\Delta}{2} \sigma_x + \omega a^{\dagger} a + g \left[ a^2 + (a^{\dagger})^2 \right] \sigma_z. \label{H_btp}
	\end{equation}
	Since both the tpQRM and a single dissipative qubit have been experimentally realized in circuit QED platforms~\cite{felicetti_two-photon_2018, naghiloo_quantum_2019}, this model could potentially be implemented by integrating the two setups. In addition, the effective two-level system of a single trapped ion—with coherent transitions and tunable dissipation—can be described by a $\mathcal{PT}$-symmetric Hamiltonian featuring balanced gain and loss~\cite{wang_observation_2021}. Given that the realization of the tpQRM in trapped-ion systems has already been proposed~\cite{felicetti_spectral_2015}, simulating the btpQRM within the same framework via laser driving appears to be experimentally feasible.
	
	The parity operator of the Hermitian tpQRM, $\Pi = \sigma_{x} \exp\left(i\frac{\pi}{2} a^{\dagger} a\right)$, is not conserved in the btpQRM, indicating that the Hamiltonian no longer exhibits $\mathbb{Z}_4$ symmetry. The time-reversal operator $\mathcal{T}$ acts by complex conjugation, satisfying $\mathcal{T} \hat{x} \mathcal{T} = \hat{x}$ and $\mathcal{T} \hat{p} \mathcal{T} = -\hat{p}$, where $\hat{x}$ and $\hat{p}$ denote the generalized position and momentum operators, respectively. As a result, we have $\mathcal{T} a (a^{\dagger}) \mathcal{T} = a (a^{\dagger})$. Therefore,
	\begin{eqnarray}
		&&\Pi \mathcal{T} H_{\mathrm{btp}} \mathcal{T} \Pi 
		= \Pi \left[ a^{\dagger} a + g\left( a^2 + (a^{\dagger})^2 \right) \sigma_{z} - i\frac{\epsilon}{2} \sigma_{z} - \frac{\Delta}{2} \sigma_{x} \right] \Pi \notag \\
		&&= a^{\dagger} a + g \left( a^2 + (a^{\dagger})^2 \right) \sigma_{z} + i\frac{\epsilon}{2} \sigma_{z} - \frac{\Delta}{2} \sigma_{x} = H_{\mathrm{btp}}.
	\end{eqnarray}
	Thus, the Hamiltonian is indeed $\mathcal{PT}$-symmetric.
	
	\textsl{Dissipative Two-Photon Quantum Rabi Model:} The Hamiltonian of the dtpQRM is given by
	\begin{equation}
		H_{\mathrm{dtp}} = \omega a^{\dagger} a + i g \left[ a^2 + (a^{\dagger})^2 \right] \sigma_z - \frac{\Delta}{2} \sigma_x. \label{H_dtp}
	\end{equation}
	The atom–field interaction described by the dtpQRM may be experimentally realized using cold atoms. A purely imaginary dissipative coupling between two long-lived atomic spin waves has already been demonstrated experimentally~\cite{peng_anti-paritytime_2016, zhang_realizing_2024}. In particular, a single cold atom in an optical cavity, driven by a transverse pump field, can interact via dissipative coupling. By tuning the pump frequency to match half the energy gap between the ground and second excited states, the first excited state is effectively eliminated from the dynamics, enabling two-photon transitions between the remaining levels.  In this work, however, we focus exclusively on theoretical analysis, aiming to provide concrete guidance for future experimental efforts.
	
	Notably, in contrast to the above btpQRM, the parity operator $\Pi$ remains conserved in the dtpQRM and satisfies $\Pi^4 = \id$, where $\id$ denotes the identity operator. This implies that the Hamiltonian possesses a $\mathbb{Z}_4$ symmetry. Consequently, the eigenfunctions of the Hamiltonian are also eigenfunctions of $\Pi$, with eigenvalues $\{1, -1, i, -i\}$. 	Meanwhile, another parity operator can be defined as $\mathcal{P} = \sigma_x \otimes \id$, under which we have
	\begin{eqnarray}
		&&\mathcal{PT} H_{\mathrm{dtp}} \mathcal{PT} = \mathcal{P} \left[ a^{\dagger} a - i g \left( a^2 + (a^{\dagger})^2 \right) \sigma_z - \frac{\Delta}{2} \sigma_x \right] \mathcal{P} \nonumber \\
		&&= a^{\dagger} a + i g \left( a^2 + (a^{\dagger})^2 \right) \sigma_z - \frac{\Delta}{2} \sigma_x = H_{\mathrm{dtp}}. 
	\end{eqnarray}
	Thus, the dtpQRM also exhibits $\mathcal{PT}$ symmetry.
	
	In the following section, we present the exact solutions of the two non-Hermitian models introduced above.
	
	\section{Exact Solutions}
	
	\subsection{Biased Two-Photon Quantum Rabi Model}
	
	The $G$-function for the btpQRM can be derived by following a procedure similar to that used for the Hermitian biased tpQRM~\cite{xie_double_2021}. We begin by applying a pair of opposite unitary transformations, defined as $S(\pm \theta) = \exp \left[ \pm i \frac{\theta}{2} \left( (a^{\dagger})^2 - a^2 \right) \right]$, to the Hamiltonian~\eqref{H_dtp}. This yields the transformed Hamiltonians $H_{\pm} = S(\pm \theta) H_{\mathrm{btp}} S(\mp \theta)$, where
	\begin{eqnarray}
		\theta = \cosh^{-1} \left( \sqrt{\frac{1 + \beta}{2\beta}} \right), \quad \beta = \sqrt{1 - 4g^2}.
	\end{eqnarray}
	
	Next, we define a set of ladder operators that satisfy the $\text{su}(1,1)$ Lie algebra:
	\begin{equation}
		K_0 = \frac{1}{2} \left( a^{\dagger} a + \frac{1}{2} \right), \quad 
		K_+ = \frac{1}{2} (a^{\dagger})^2, \quad 
		K_- = \frac{1}{2} a^2,
	\end{equation}
	with the commutation relations
	\begin{equation}
		[K_0, K_{\pm}] = \pm K_{\pm}, \quad 
		[K_+, K_-] = -2K_0.
	\end{equation}
	The Hilbert space $\mathcal{H}$, generated by the action of $a^{\dagger}$ on the photon vacuum state $\vert 0 \rangle$, decomposes into two irreducible subspaces characterized by the Bargmann index $q$, defined via $K_0 \vert q, 0 \rangle = q \vert q, 0 \rangle$. The even-photon-number subspace is given by $\mathcal{H}_{\frac{1}{4}} = \left\{ (a^{\dagger})^n \vert 0 \rangle \mid n = 0, 2, 4, \ldots \right\}$ with $q = \frac{1}{4}$, while the odd-photon-number subspace is $\mathcal{H}_{\frac{3}{4}} = \left\{ (a^{\dagger})^n \vert 0 \rangle \mid n = 1, 3, 5, \ldots \right\}$ with $q = \frac{3}{4}$. The basis states $\vert q, n \rangle$ are explicitly given by
	\begin{eqnarray}
		\vert q, n \rangle &=& \left\vert 2\left( q + n - \frac{1}{4} \right) \right\rangle 
		= \frac{(a^{\dagger})^{2(q + n - \frac{1}{4})}}{\sqrt{[2(q + n - \frac{1}{4})]!}} \vert 0 \rangle, \\
		K_0 \vert q, n \rangle &=& (q + n) \vert q, n \rangle.
	\end{eqnarray}
	
	In terms of the $\{K_0, K_{\pm}\}$ operators, the Hamiltonians $H_{\pm}$ can be rewritten as
	\begin{eqnarray}
		&&H_{+}^{(K)} =
		\nonumber \\
		&&\begin{bmatrix}
			2 \beta K_0 + i \dfrac{\epsilon}{2} & - \dfrac{\Delta}{2} \\
			- \dfrac{\Delta}{2} & \dfrac{2K_0 (1 + 4g^2) - 4g (K_+ + K_-)}{\beta} - i \dfrac{\epsilon}{2}
		\end{bmatrix} - \dfrac{\id}{2}, \nonumber \\
		&&H_{-}^{(K)} =
		\nonumber \\
		&&\begin{bmatrix}
			\dfrac{2K_0 (1 + 4g^2) + 4g (K_+ + K_-)}{\beta} + i \dfrac{\epsilon}{2} & - \dfrac{\Delta}{2} \\
			- \dfrac{\Delta}{2} & 2 \beta K_0 - i \dfrac{\epsilon}{2}
		\end{bmatrix} - \dfrac{\id}{2}. \nonumber \\
	\end{eqnarray}
	We now propose a general expansion for the eigenfunctions of $H_{\pm}^{(K)}$ as
	\begin{equation}
		\left\vert \psi^{(q)}_{\pm} \right\rangle =
		\begin{bmatrix}
			\sum\limits_{n=0}^{\infty} \sqrt{ \left[ 2 \left( n + q - \frac{1}{4} \right) \right]! } \; e_{n,\pm}^{(q)} \left\vert q, n \right\rangle \\
			\sum\limits_{n=0}^{\infty} \sqrt{ \left[ 2 \left( n + q - \frac{1}{4} \right) \right]! } \; f_{n,\pm}^{(q)} \left\vert q, n \right\rangle
		\end{bmatrix},
		\label{psi_K_pm}
	\end{equation}
	where $e_{n,\pm}^{(q)}$ and $f_{n,\pm}^{(q)}$ are expansion coefficients. Substituting into the Schrödinger equation $H_{\pm}^{(K)} \left\vert \psi_{\pm}^{(q)} \right\rangle = E \left\vert \psi_{\pm}^{(q)} \right\rangle$ and projecting onto the basis states $\left\vert q, n \right\rangle$, we obtain the following recurrence relations:
	\begin{subequations}
		\label{recur_btp}
		\begin{equation}
			e_{n,\pm}^{(q)} = \frac{ \dfrac{\Delta}{2} f_{n,\pm}^{(q)} }{ 2(n + q)\beta \pm i \dfrac{\epsilon}{2} - \dfrac{1}{2} - E },
			\label{recur_btp_en}
		\end{equation}
		\begin{eqnarray}
			f_{n+1,\pm}^{(q)} &=& \frac{ \left[ 2(n + q)(1 + 4g^2) - \beta \left( \pm i \dfrac{\epsilon}{2} + \dfrac{1}{2} + E \right) \right] f_{n,\pm}^{(q)} }{ 8g (n + q + \frac{1}{4})(n + q + \frac{3}{4}) } \nonumber \\
			&& - \frac{ 2g f_{n-1,\pm}^{(q)} + \dfrac{\Delta}{2} \beta e_{n,\pm}^{(q)} }{ 8g (n + q + \frac{1}{4})(n + q + \frac{3}{4}) }.
			\label{recur_btp_fn}
		\end{eqnarray}
	\end{subequations}
	
	Transforming back to the original Hamiltonian, the eigenfunctions are given by $\vert \Psi_{\pm}^{(q)} \rangle = S(\mp \theta) \vert \psi_{\pm}^{(q)} \rangle$. Since both expressions represent the same physical state, the two must be proportional, i.e., $\vert \Psi_{+}^{(q)} \rangle \propto \vert \Psi_{-}^{(q)} \rangle$. By projecting both states onto the vacuum state $\vert q, 0 \rangle$ and using 
	\begin{equation}
		\langle q, 0 \vert S(\theta) \vert q, n \rangle \propto \frac{ \sqrt{ \left[ 2\left( n + q - \frac{1}{4} \right) \right]! } }{n!} \left( \frac{\tanh \theta}{2} \right)^n,
		\label{projct_0_btpQRM}
	\end{equation}
	the $G$-function can be formulated as
	\begin{eqnarray}
		G^{(q)} &=& \left( \sum_{n=0}^{\infty} F(n) \, e_{n,+}^{(q)} \right) \left( \sum_{n=0}^{\infty} F(n) \, e_{n,-}^{(q)} \right) \nonumber \\
		&& - \left( \sum_{n=0}^{\infty} F(n) \, f_{n,+}^{(q)} \right) \left( \sum_{n=0}^{\infty} F(n) \, f_{n,-}^{(q)} \right),
		\label{G_btpQRM}
	\end{eqnarray}
	where the weight function $F(n)$ is defined as
	\begin{equation}
		F(n) = \frac{ \left[ 2\left( n + q - \frac{1}{4} \right) \right]! }{n!} \left( \frac{\tanh \theta}{2} \right)^n.
	\end{equation}
	The coefficients $e_{n,\pm}^{(q)}$ and $f_{n,\pm}^{(q)}$ are determined by the recurrence relations in Eq.~\eqref{recur_btp}, with the initial condition $f_{0,\pm}^{(q)} \equiv 1$. The zeros of the $G$-function correspond to the eigenvalue spectrum.
	
	\begin{figure}[tbp]
		\centering
		\subfloat[\label{realG_btpQRM_fig}]{\includegraphics[width=1.0\linewidth]{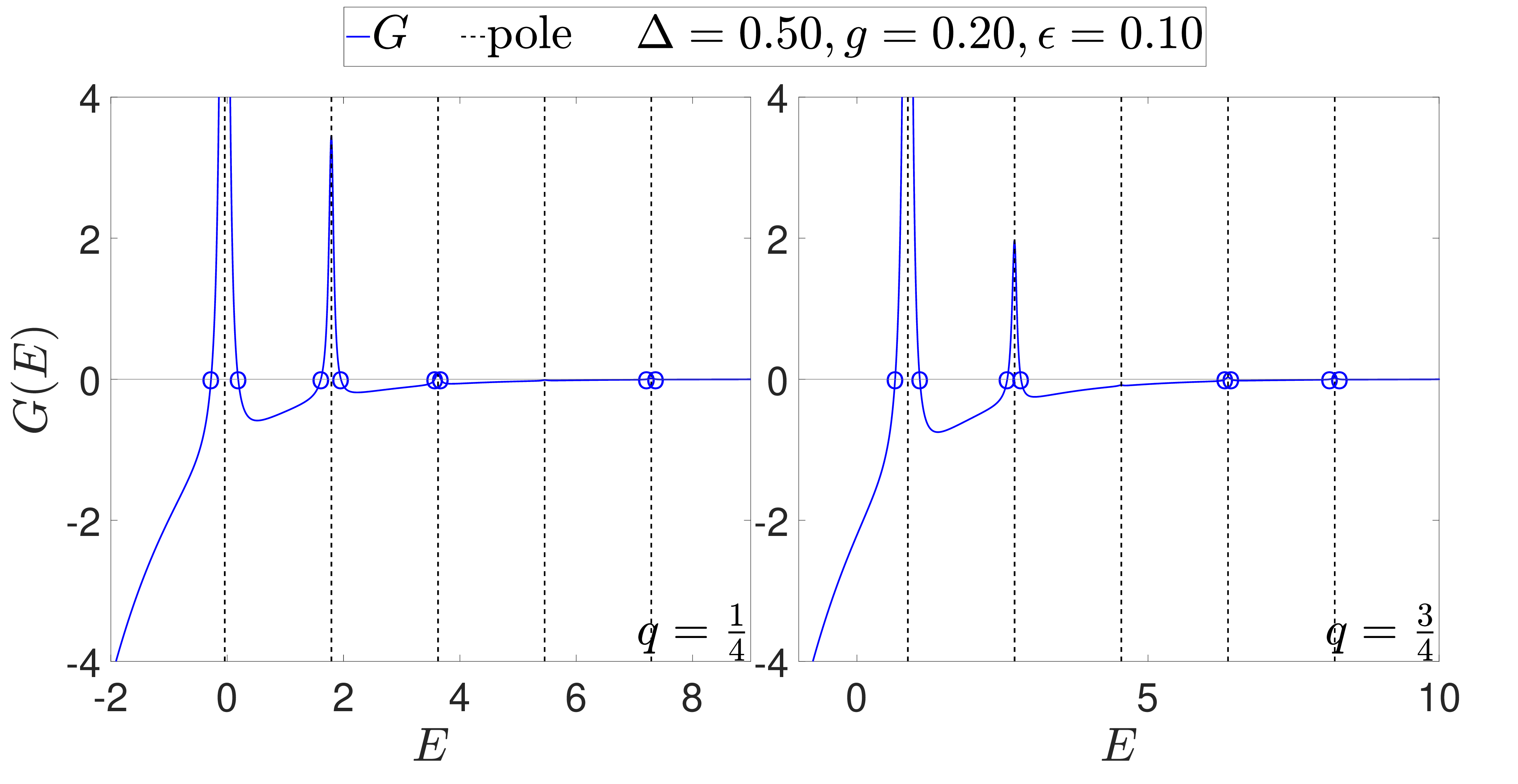}}
		\newline
		\subfloat[\label{fullG_btp_fig}]{\includegraphics[width=1.0\linewidth]{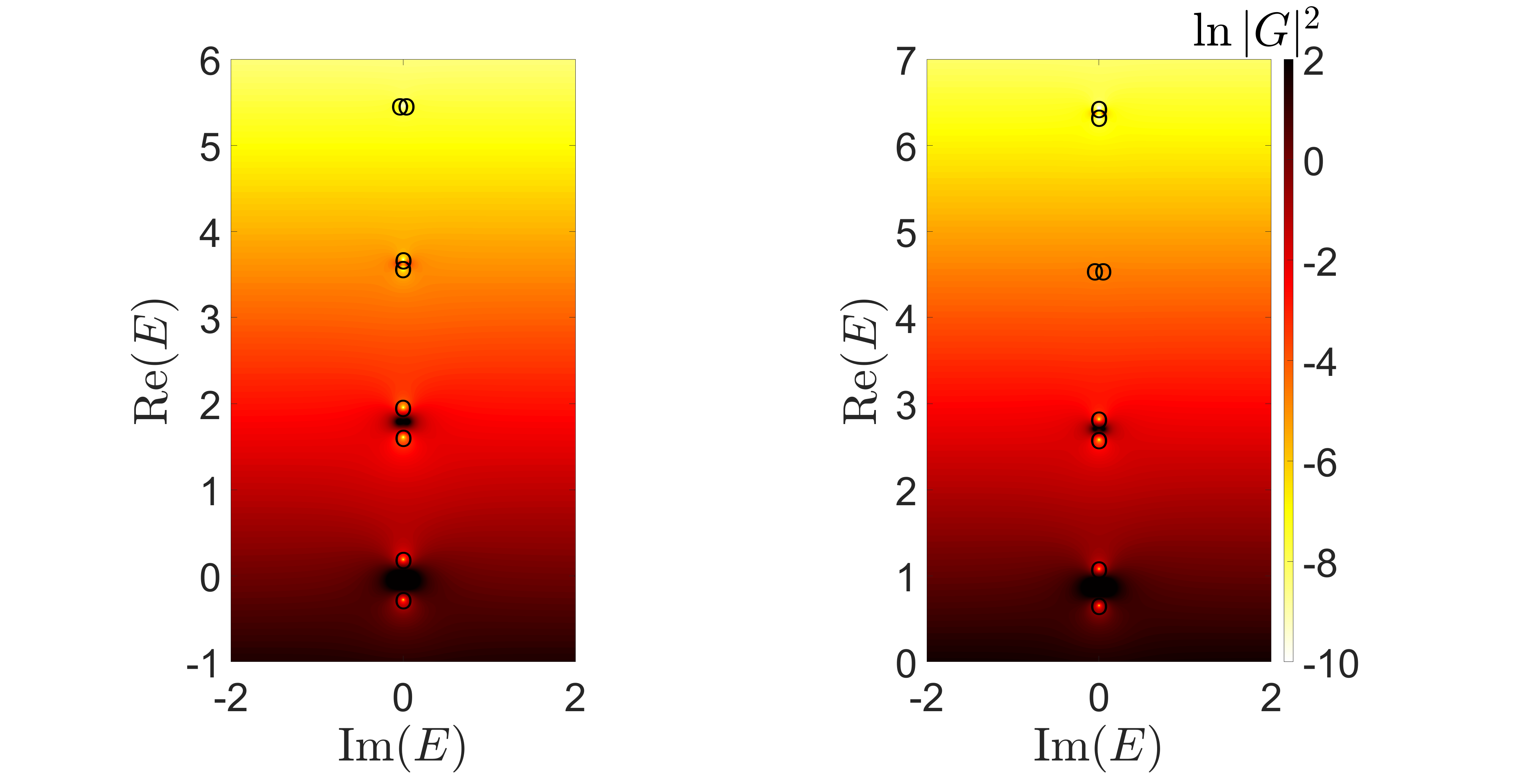}}
		\caption{(a) $G$-function curves of the btpQRM in the real energy regime for $q = 1/4$ (left) and $q = 3/4$ (right). Blue lines represent the $G$-function, while black dashed lines indicate the pole positions $E_{n,0}^{(q,\mathrm{pole})}$. (b) Distribution of $\ln |G|^2$ in the complex energy plane for $q = 1/4$ (left) and $q = 3/4$ (right). In both panels, $\Delta = 0.50$, $g = 0.20$, and the open circles mark the zeros of the $G$-function.}
		\label{G_btpQRM_fig}
	\end{figure}
	
	Since $E$ and $i\epsilon$ always appear together in the recurrence relations \eqref{recur_btp}, it follows that $e_{n,+}^{(q) *}(E) = e_{n,-}^{(q)}(E^*)$ and $f_{n,+}^{(q) *}(E) = f_{n,-}^{(q)}(E^*)$. Consequently, the $G$-function satisfies $G^{(q)}(E^*) = \left[ G^{(q)}(E) \right]^*$, implying that if $E$ is a solution, then its complex conjugate $E^*$ must also be a solution. This confirms that the spectrum is conjugate symmetric. Furthermore, in the $\mathcal{PT}$-symmetric regime where $E$ is real, the expansion coefficients satisfy $e_{n,+}^{(q) *}(E) = e_{n,-}^{(q)}(E)$ and $f_{n,+}^{(q) *}(E) = f_{n,-}^{(q)}(E)$, and the $G$-function simplifies to
	\begin{equation}
		G_{\mathcal{PT}}^{(q)} = \left| \sum_{n=0}^{\infty} F(n) \, e_{n,+}^{(q)} \right|^2 - \left| \sum_{n=0}^{\infty} F(n) \, f_{n,+}^{(q)} \right|^2,
	\end{equation}
	which is manifestly real.
	
	The zeros of $G_{\mathcal{PT}}^{(q)}$, corresponding to real eigenvalues, are generally located near the pole lines, as shown in Fig.~\ref{realG_btpQRM_fig}. However, the absence of zeros in the vicinity of certain pole lines signals the emergence of complex eigenvalues and the breaking of $\mathcal{PT}$ symmetry. As illustrated in Fig.~\ref{fullG_btp_fig}, both real and complex zeros of the $G$-function—indicated by open circles—exhibit a symmetric distribution about the imaginary axis.
	
	\subsection{Dissipative Two-Photon Quantum Rabi Model}
	
	The $G$-function for the dtpQRM can be derived using a procedure analogous to the Hermitian tpQRM~\cite{duan_two-photon_2016}.  The key difference lies in the use of a similarity transformation that preserves the eigenvalue spectrum:
	\begin{equation}
		S(ir) = \exp \left[ i \frac{r}{2} \left( (a^{\dagger})^2 - a^2 \right) \right],
	\end{equation}
	where $2r = \cos^{-1}(1/\gamma)$ and $\gamma = \sqrt{1 + 4g^2}$. This transformation yields the following operator identities:
	\begin{eqnarray}
		&&S(ir) \, a \, S(-ir) = a \cos r - i a^{\dagger} \sin r, \nonumber \\
		&&S(ir) \, a^{\dagger} \, S(-ir) = a^{\dagger} \cos r - i a \sin r, \nonumber \\
		&&S(ir) S(-ir) = \id, \quad S(ir)^{\dagger} S(ir) = S(2ir) \neq \id.
	\end{eqnarray}

	Applying the similarity transformation to the Hamiltonian \eqref{H_dtp} yields the transformed Hamiltonian $H_S = S(ir) H_{\mathrm{dtp}} S(-ir)$. In terms of $\{K_0, K_{\pm}\}$, $H_S$ can be written as
	\begin{equation}
		H_S^{(K)} =
		\begin{bmatrix}
			2\gamma K_0 & - \frac{\Delta}{2} \\
			- \frac{\Delta}{2} & \dfrac{2K_0(2 - \gamma^2) - 4i g (K_+ + K_-)}{\gamma}
		\end{bmatrix} - \frac{\id}{2}.
	\end{equation}
	The eigenfunction of $H_S^{(K)}$ is expressed as
	\begin{equation}
		\left\vert \psi^{(q)} \right\rangle =
		\begin{bmatrix}
			\sum\limits_{n=0}^{\infty} \sqrt{ \left[ 2\left( n + q - \frac{1}{4} \right) \right]! } \, i^{-n} e_n^{(q)} \left\vert q, n \right\rangle \\
			\sum\limits_{n=0}^{\infty} \sqrt{ \left[ 2\left( n + q - \frac{1}{4} \right) \right]! } \, i^{-n} f_n^{(q)} \left\vert q, n \right\rangle
		\end{bmatrix},
		\label{psi_K_dtpQRM}
	\end{equation}
	where the expansion coefficients $e_n^{(q)}$ and $f_n^{(q)}$ satisfy the following recurrence relations:
	\begin{subequations}
		\label{recur_dtpQRM}
		\begin{equation}
			e_n^{(q)} = \frac{ \frac{\Delta}{2} f_n^{(q)} }{ 2(n + q)\gamma - \frac{1}{2} - E },
			\label{recur_dtpQRM_en}
		\end{equation}
		\begin{eqnarray}
			f_{n+1}^{(q)} &=& \frac{ \left[ 2(n + q)(2 - \gamma^2) - \gamma \left( \frac{1}{2} + E \right) \right] f_n^{(q)} }{ 8g (n + q + \frac{1}{4})(n + q + \frac{3}{4}) } \nonumber \\
			&& + \frac{ 2g f_{n-1}^{(q)} - \frac{\Delta}{2} \gamma e_n^{(q)} }{ 8g (n + q + \frac{1}{4})(n + q + \frac{3}{4}) },
			\label{recur_dtpQRM_fn}
		\end{eqnarray}
	\end{subequations}
	which can be obtained as privious. 
	
	Transforming back to the original Hamiltonian, the eigenfunction is given by $\vert \Psi^{(q)} \rangle = S(-ir) \vert \psi_{\mathrm{dtp}}^{(q)} \rangle$. As previously discussed, the parity operator $\Pi$ is conserved in the dtpQRM, implying that $\Pi \vert \Psi^{(q)} \rangle \propto \vert \Psi^{(q)} \rangle$. By projecting onto the corresponding vacuum state $\vert q, 0 \rangle$ and using 
	\begin{equation}
		\langle q, 0 \vert S(-ir) \vert q, n \rangle \propto \frac{ \sqrt{ \left[ 2 \left( n + q - \frac{1}{4} \right) \right]! } }{n!} \left( \frac{i \tan r}{2} \right)^n,
		\label{projct_0_dtpQRM}
	\end{equation}
	the $G$-function is obtained as
	\begin{equation}
		G_{\pm}^{(q)} = \sum_{n=0}^{\infty} \left( e_n^{(q)} \mp f_n^{(q)} \right) \frac{ \left[ 2 \left( n + q - \frac{1}{4} \right) \right]! }{n!} \left( \frac{\tan r}{2} \right)^n,
		\label{G_dtpQRM}
	\end{equation}
	where the subscripts $\pm$ correspond to even and odd parity under $\Pi$, respectively. The coefficients $e_n^{(q)}$ and $f_n^{(q)}$ are determined by the recurrence relations~\eqref{recur_dtpQRM}, with the initial condition $f_0^{(q)} \equiv 1$. The zeros of the $G$-function yield the regular eigenvalue spectrum.
	
	\begin{figure}[tbp]
		\centering
		\subfloat[\label{realG_dtpQRM_fig}]{\includegraphics[width=1.0\linewidth]{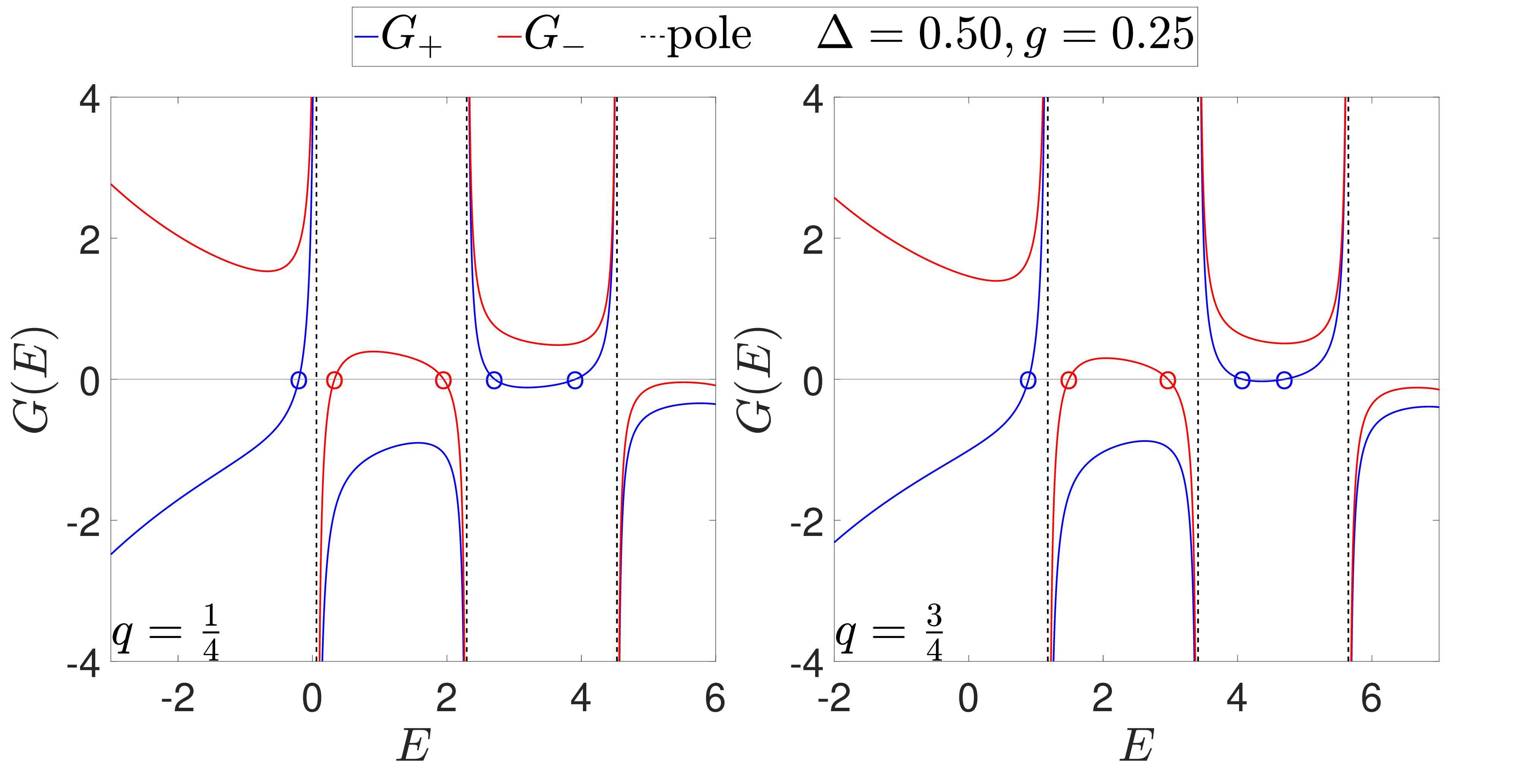}}
		\newline
		\subfloat[\label{fullG_dtp_fig}]{\includegraphics[width=1.0\linewidth]{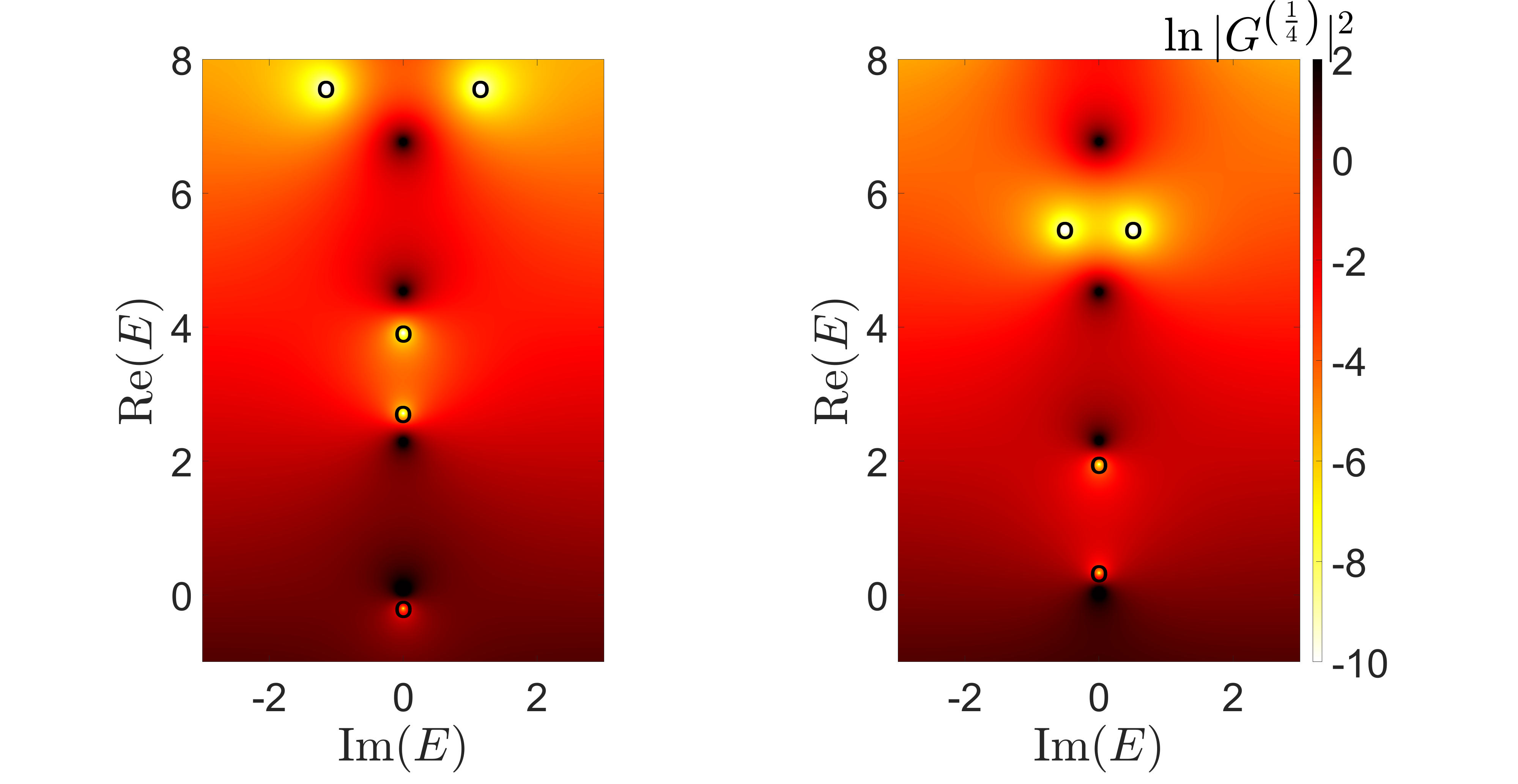}}
		\caption{(a) $G$-function curves of the dtpQRM in the real energy regime for $q = 1/4$ (left) and $q = 3/4$ (right). Blue (red) lines represent the $G_{+}$ ($G_{-}$) functions. Black dashed lines indicate the pole positions $E_{n}^{(q,\mathrm{pole})}$. (b) Distribution of $\ln |G_{+}|^2$ (left) and $\ln |G_{-}|^2$ (right) in the complex energy plane for $q = 1/4$. In both panels, $\Delta = 0.50$, $g = 0.25$, and open circles mark the zeros of the $G$-function.}
		\label{G_dtpQRM_fig}
	\end{figure}
	
	Since the energy $E$ is the only parameter that can take complex values in the recurrence relations~\eqref{recur_dtpQRM}, the $G$-function satisfies the conjugation symmetry $G_{\pm}^{(q)}(E^*) = \left[ G_{\pm}^{(q)}(E) \right]^*$. This implies that if $E$ is a zero of the $G$-function, then its complex conjugate $E^*$ is also a zero, confirming that the spectrum is conjugate symmetric. When $E$ is real, the $G$-function remains purely real. In the high-energy regime, however, the real zeros gradually vanish, as shown in Fig.~\ref{realG_dtpQRM_fig}, indicating the emergence of complex eigenvalues and the onset of $\mathcal{PT}$-symmetry breaking. As illustrated in Fig.~\ref{fullG_dtp_fig}, both real and complex zeros—indicated by open circles—are symmetrically distributed with respect to the imaginary axis in the complex energy plane.
	
	\section{Exceptional Point and Doubly Degenerate Point}
	
	\subsection{Biased Two-Photon Quantum Rabi Model}
	
	\textsl{Spectral Collapse:} We begin by examining whether the hallmark feature of spectral collapse in the Hermitian tpQRM persists in the btpQRM. As demonstrated by \textcite{braak_spectral_2023}, the pole structure of the $G$-function determines the collapse point. This structure is governed by the condition under which the denominator of $e_{n,\pm}^{(q)}$~\eqref{recur_btp_en} vanishes:
	\begin{equation}
		E_{n,\pm}^{(q,\mathrm{pole})} = 2(n + q)\beta \pm i \frac{\epsilon}{2} - \frac{1}{2}.
	\end{equation}
	
	When $\epsilon = 0$, the poles reduce to $E_{n,\pm}^{(q,\mathrm{pole})} = E_{n,0}^{(q,\mathrm{pole})} = 2(n + q)\beta - \frac{1}{2}$, which are purely real and degenerate. This recovers the exact pole structure of the $G$-function in the Hermitian tpQRM. In this case, all pole lines collapse to a single point at the critical coupling strength $g_c = 1/2$, where $\beta = 0$ and the Bogoliubov transformation becomes singular. Consequently, spectral collapse occurs at $g = g_c$, and the energy spectrum becomes continuous above the threshold energy $E_c = -1/2$.
	
	\begin{figure}[tbp]
		\centering
		\includegraphics[width=1.0\linewidth]{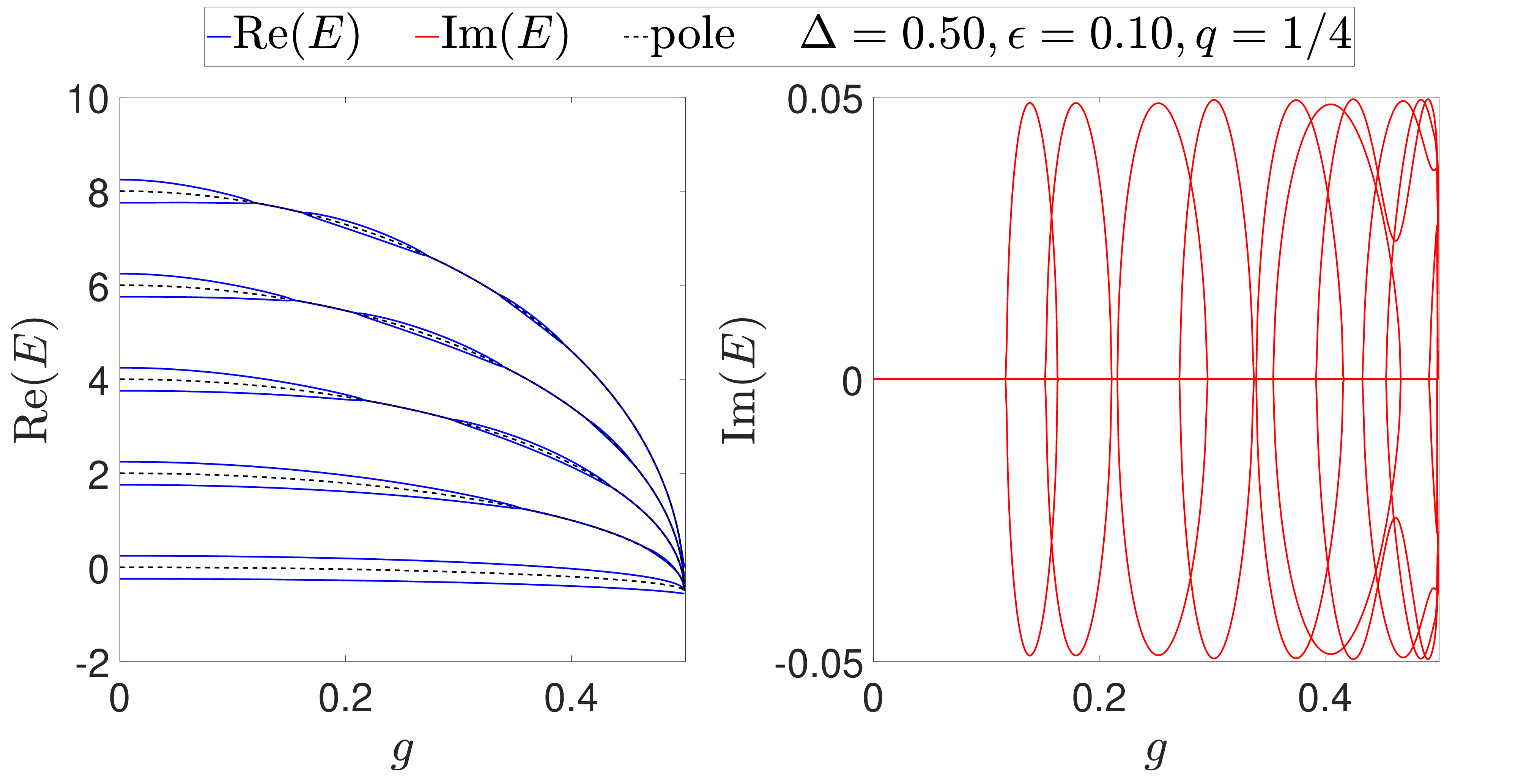}
		\caption{Real (left) and imaginary (right) parts of the lowest few eigenvalues of the btpQRM as a function of the coupling strength $g$. Dashed lines indicate the pole positions $E_{n,0}^{(q,\mathrm{pole})}$. Here, $\Delta = 0.50$, $\epsilon = 0.10$, and $q = 1/4$.}
		\label{sepctra_bias_delta0.50_fig}
	\end{figure}
	
	We observe that when $g = g_c$ and $\epsilon \neq 0$, the real parts of all pole lines still merge at a single point $E = E_c$, and the Bogoliubov transformation becomes singular. This indicates that spectral collapse persists at $g = g_c$. As shown in Fig.~\ref{sepctra_bias_delta0.50_fig}, the real part of the spectrum indeed collapses at $g = g_c$ and $E = E_c$, while the imaginary part spreads over the interval $[-\epsilon/2, \epsilon/2]$.
	
	This behavior becomes more transparent in the $x$-representation. At the collapse point $g = g_c$, the Hamiltonian~\eqref{H_btp} takes the form
	\begin{equation}
		H_\mathrm{btp}^{(x)} = \begin{bmatrix}
			x^2 + i \dfrac{\epsilon}{2} - \dfrac{1}{2} & -\dfrac{\Delta}{2} \\
			-\dfrac{\Delta}{2} & p^2 - i \dfrac{\epsilon}{2} - \dfrac{1}{2}
		\end{bmatrix}.
	\end{equation}
	It is evident that when $\Delta = 0$, the qubit decouples from the field. In this case, the real part of the spectrum becomes fully continuous above the threshold energy $E_c = -1/2$, and the imaginary parts are fixed at $\pm \epsilon/2$. When $\Delta \neq 0$, the spectral collapse is incomplete: discrete bound states remain below $E_c$, and the imaginary parts of the spectrum are distributed across the interval $[-\epsilon/2, \epsilon/2]$.
	
	\begin{figure}[tbp]
		\centering
		\subfloat[\label{sepctra_bias_delta0.50_h_fig}]{\includegraphics[width=1.0\linewidth]{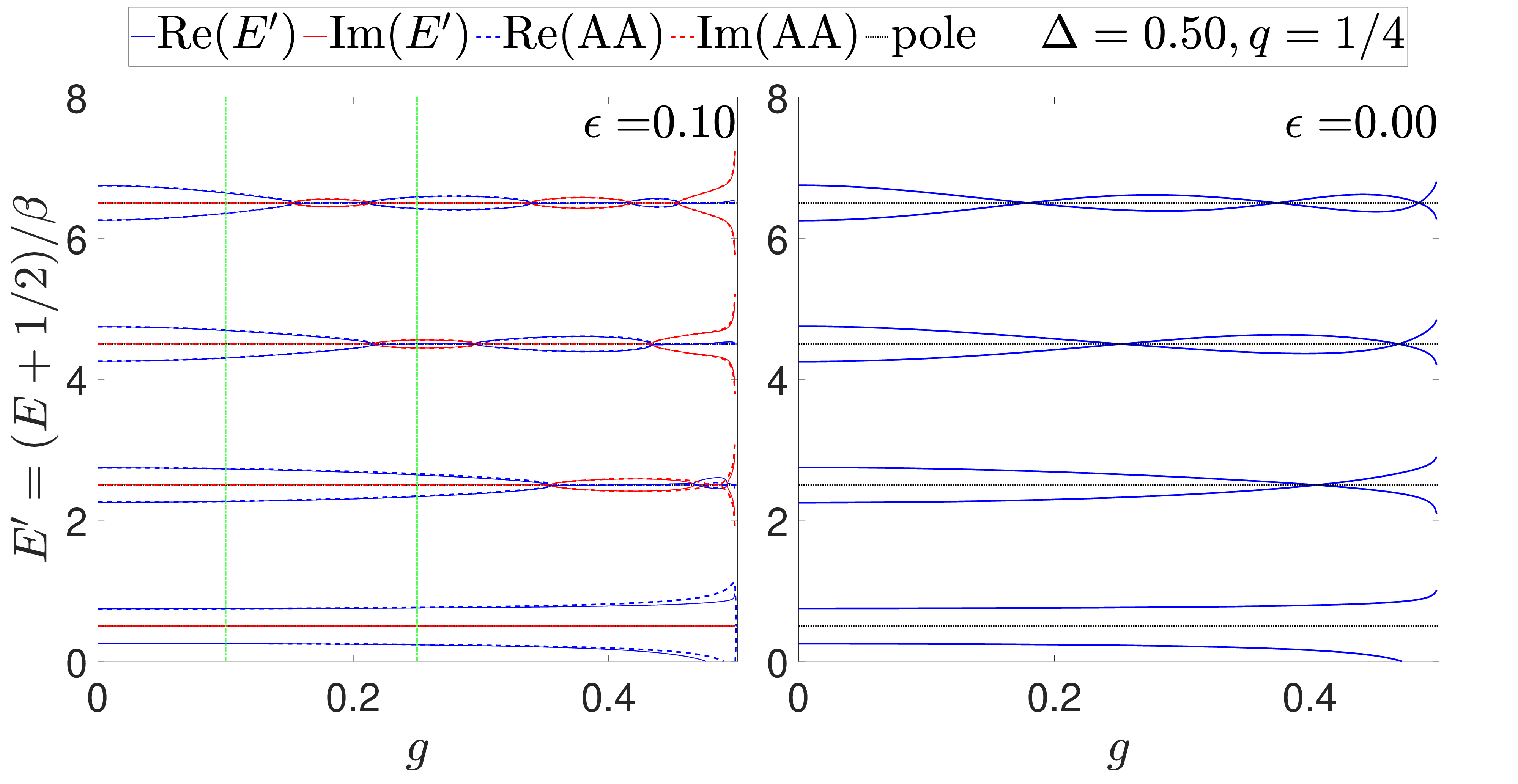}}
		\newline
		\subfloat[\label{sepctra_bias_delta0.50ep0.40_fig}]{\includegraphics[width=1.0\linewidth]{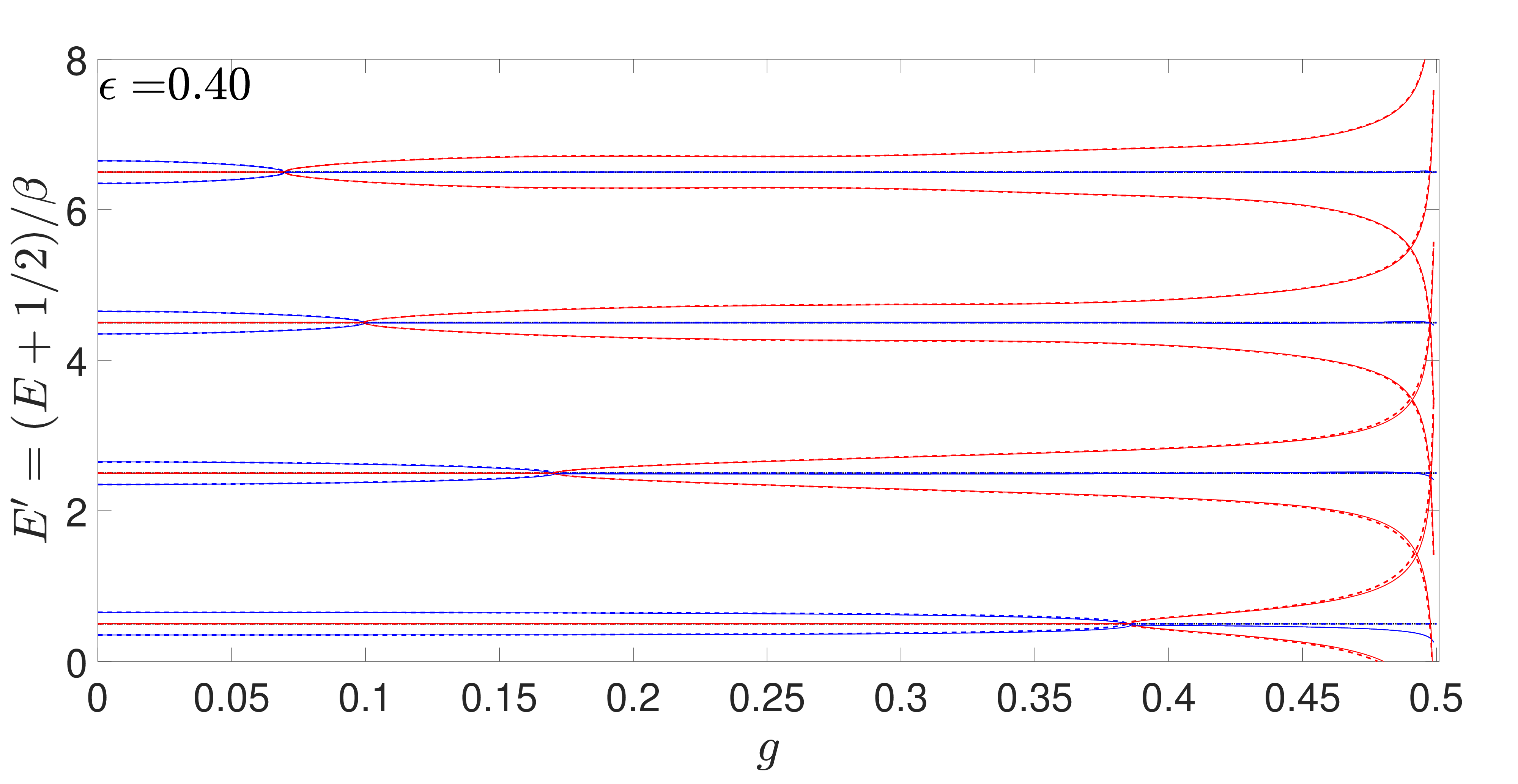}}
		\caption{Scaled spectra $E' = (E + 1/2)/\beta$ at $\Delta = 0.50$ and $q = 1/4$. (a) The comparison between the btpQRM (left, $\epsilon = 0.10$) and the Hermitian tpQRM (right, $\epsilon = 0.00$). (b) is scaled spectra for the btpQRM at $\epsilon=0.40$. Blue (red) lines represent the real (imaginary) part of $E'$, while blue (red) dashed lines show the corresponding results obtained from the adiabatic approximation. Black dotted lines indicate $E_{n,0}^{(q,\mathrm{pole})}$. The green lines mark the parameters used in Sec.~\ref{sec_dyna_btp}}
		
	\end{figure}
	
	\textsl{Exceptional Points:} It can be seen that the $\mathcal{PT}$-broken regions vanish and revive along the pole lines $E_{n,0}^{(q,\mathrm{pole})}$, with their endpoints corresponding to exceptional points EPs. This pattern naturally evokes the doubly degenerate points in the Hermitian tpQRM, known as Juddian solutions~\cite{emary_bogoliubov_2002}, where eigenstates of different parity $\Pi$ intersect along the pole lines. By comparing the spectra of the btpQRM and the Hermitian tpQRM in Fig.~\ref{sepctra_bias_delta0.50_h_fig}, we observe that the $\mathcal{PT}$-broken regions in the btpQRM correspond to the Juddian degeneracies in the Hermitian case. Moreover, the number of $\mathcal{PT}$-broken segments along each pole line increases with the energy level index. This correspondence arises because the parity operator $\Pi$ is no longer conserved in the btpQRM—an effect also observed in the one-photon $\mathcal{PT}$-symmetric QRM~\cite{lu_pt_2023}.
	
	This correspondence allows us to analytically determine the locations of $\mathcal{PT}$-broken regions by examining the case $\epsilon = 0$. In this limit, we have $e_{n,+}^{(q)} = e_{n,-}^{(q)}$ and $f_{n,+}^{(q)} = f_{n,-}^{(q)}$. At $E = E_{n,0}^{(q,\mathrm{pole})}$, $e_{n,\pm}^{(q)}$ remains well-defined only if $f_{n,\pm}^{(q)} = 0$. By iterating the recurrence relations~\eqref{recur_btp}, one can thus identify the locations of the doubly degenerate points along the $n$th pole line. For instance, the degeneracy point on the first pole line is given by
	\begin{eqnarray}
		g_{1,0}^{(1/4)} = \frac{1}{4} \sqrt{\frac{16 - \Delta^2}{6}}, \nonumber \\
		E_{1,0}^{(1/4)} = \frac{5}{2} \sqrt{ \frac{8 + \Delta^2}{24} } - \frac{1}{2}.
	\end{eqnarray}
	When $\Delta > 4$, no doubly degenerate points exist on the first pole line, indicating that the corresponding levels remain $\mathcal{PT}$-symmetric in the btpQRM. Moreover, since $f_{0,\pm}^{(q)} \equiv 1$, the first-order quantum phase transition does not occur in the Hermitian tpQRM. As a result, the ground state and the first excited state in the btpQRM are always $\mathcal{PT}$-symmetric.
	
	It is important to emphasize that the above conclusions hold primarily when the bias strength $\epsilon$ is small compared to $\Delta$. The adiabatic approximation (AA) provides an analytically tractable approach~\cite{duan_two-photon_2016}. In the basis $[S(-\theta) \vert n \rangle, S(\theta) \vert n \rangle]$, and under the assumption that tunneling occurs only between states with the same quantum number $n$, the Hamiltonian~\eqref{H_btp} reduces to a block-diagonal form:
	\begin{equation}
		H_n^{(q)} = 2(n + q)\beta - \frac{1}{2} + i \frac{\epsilon}{2} \sigma_z - \frac{D_n^{(q)}}{2} \sigma_x,
	\end{equation}
	where
	\begin{equation}
		D_n^{(q)}(\beta) = \Delta \, \beta^{1/2} \, P_{2n + 2(q - \frac{1}{4})}(\beta),
	\end{equation}
	and $P_m(\beta)$ denotes the Legendre polynomial. The corresponding eigenvalues are given by
	\begin{equation}
		E_{n,\pm}^{(q)} = 2(n + q)\beta - \frac{1}{2} \pm \frac{1}{2} \sqrt{ \left( D_n^{(q)} \right)^2 - \epsilon^2 }.
	\end{equation}
	
	It follows that $\mathcal{PT}$ symmetry is broken when $|D_n^{(q)}| < \epsilon$. Since $|P_m(\beta)| \leq 1$, the AA predicts that all levels undergo $\mathcal{PT}$-symmetry breaking when
	\begin{equation}
		g > g_\mathrm{PTB}^{(0)} = \frac{1}{2} \sqrt{ 1 - \left( \frac{\epsilon}{\Delta} \right)^4 }.
	\end{equation}
	This critical threshold can be lower for higher-eigenvalue levels, given that the maximum of the Legendre polynomial remains bounded by 1. The AA result shows excellent agreement with the exact solutions and explains why, at the collapse point in Fig.~\ref{sepctra_bias_delta0.50_h_fig}, all levels except the ground state and first excited state exhibit $\mathcal{PT}$-symmetry breaking. For larger values of $\epsilon$, this kind of $\mathcal{PT}$-broken region may emerge even prior to the onset of the corresponding Juddian solutions, thus $\mathcal{PT}$-symmetry cannot be retained. As illustrated in Fig.~\ref{sepctra_bias_delta0.50ep0.40_fig}, even the ground state becomes $\mathcal{PT}$-broken when $g > g_\mathrm{PTB}^{(0)}$ at $\epsilon = 0.40$. In fact, when $\epsilon > \Delta$, all eigenstates exhibit $\mathcal{PT}$-symmetry breaking, as the qubit subsystem itself ceases to be $\mathcal{PT}$-symmetric.
	
	The calculation of fidelity susceptibility is a well-established method for detecting phase transitions in Hermitian systems. This approach can be generalized to the non-Hermitian regime using a biorthogonal formalism, as demonstrated by \textcite{tzeng_hunting_2021}. Fidelity, which quantifies the similarity between two quantum states, is extended in the biorthogonal basis as
	\begin{equation}
		\mathcal{F} = \langle L(\lambda) \vert R(\lambda + \delta) \rangle \langle L(\lambda + \delta) \vert R(\lambda) \rangle,
	\end{equation}
	where $\vert L \rangle$ and $\vert R \rangle$ denote the bra and ket of the biorthogonal basis, $\lambda$ is a tunable system parameter, and $\delta$ is a small perturbation. The normalization conditions are chosen such that $\langle L \vert R \rangle = 1$ and $\langle L \vert L \rangle = \langle R \vert R \rangle$. The fidelity susceptibility is then defined as
	\begin{equation}
		\chi = \frac{1 - \mathcal{F}}{\delta^2},
		\label{fidelity_sus}
	\end{equation}
	and its real part, $\mathrm{Re}(\chi)$, diverges negatively as the coupling strength $g$ approaches an EP.
	
	\begin{figure}[tbp]
		\centering
		\includegraphics[width=1.0\linewidth]{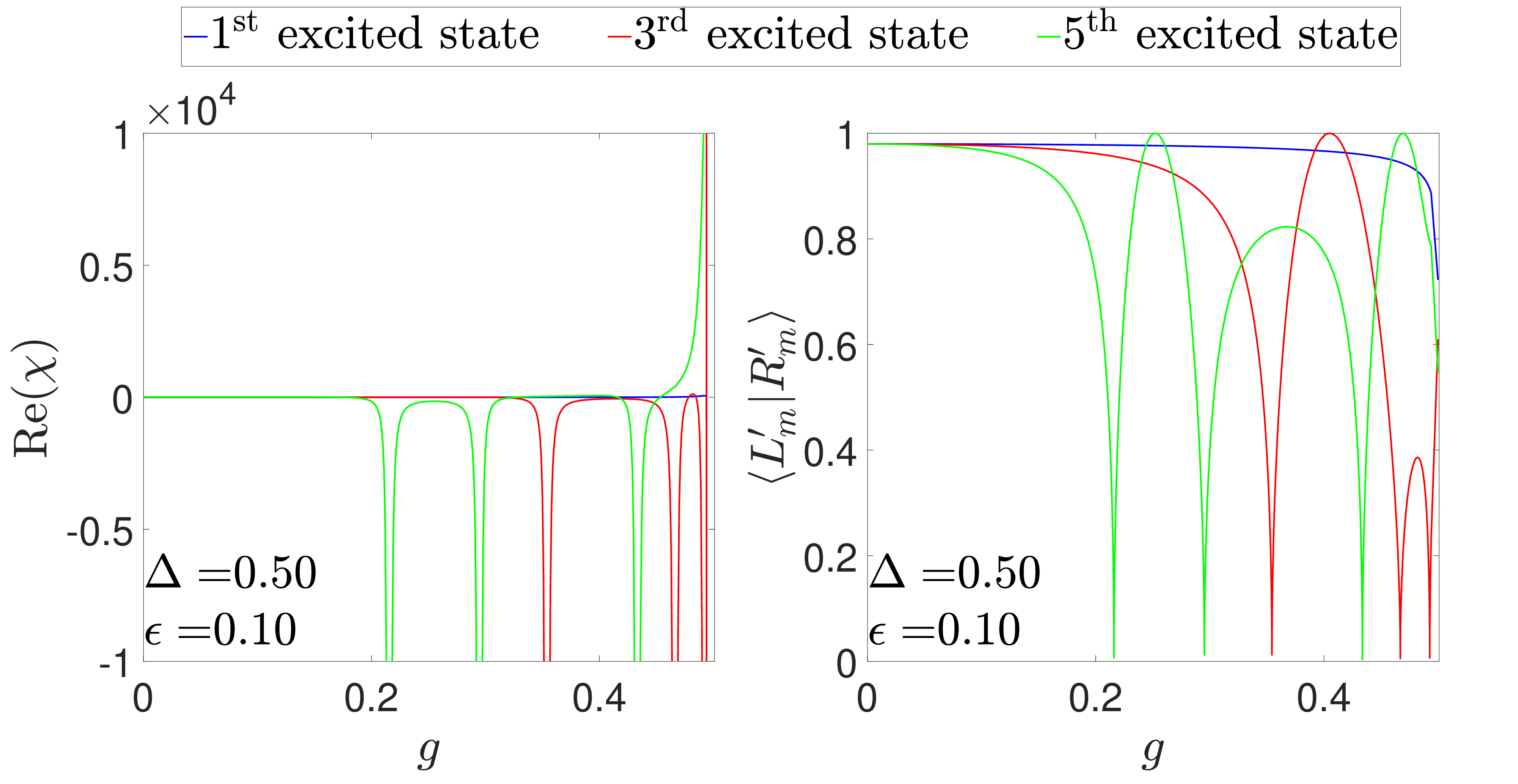}
		\caption{The real part of the fidelity susceptibility (left) and the c-product $\langle L' \vert R' \rangle$ (right) for the btpQRM, plotted as functions of the coupling strength $g$ using the same parameters as in Fig.~\ref{sepctra_bias_delta0.50_fig}. Blue, red, and green lines correspond to the $1^{\text{st}}$, $3^{\text{rd}}$, and $5^{\text{th}}$ excited states, respectively, with $q = 1/4$, $\Delta = 0.50$, and $\epsilon = 0.10$.}
		\label{fidelitysus_btp_fig}
	\end{figure}
	
	As illustrated in Fig.~\ref{fidelitysus_btp_fig}, the real part of the fidelity susceptibility, $\mathrm{Re}(\chi)$, serves as an effective indicator of EPs. Specifically, $\mathrm{Re}(\chi)$ becomes markedly large and negative in the vicinity of EPs. In practical numerical calculations, this divergence is limited by the finite perturbation strength $\delta$ and thus does not reach true infinity. Ideally, however, $\mathrm{Re}(\chi)$ diverges to $-\infty$ at EPs. Away from EPs, the eigenfunctions evolve smoothly as the coupling strength $g$ varies. However, the fidelity $\mathcal{F}$ may exceed the conventional range $[0, 1]$, since the norms $\langle L \vert L \rangle$ and $\langle R \vert R \rangle$ generally exceed 1 for complex eigenvalues. At EPs, due to self-orthogonality, it is common to observe $|\mathcal{F}| > 1$ near EPs, which explains the negative divergence of the fidelity susceptibility~\cite{tzeng_hunting_2021}.
	
	In contrast, $\mathrm{Re}(\chi)$ exhibits a pronounced positive peak as $g$ approaches the spectral collapse point. This behavior arises from the near-degeneracy of eigenvalue levels in that region, resembling the response observed near Juddian degeneracies. A more detailed analysis of this phenomenon will be presented in the context of the dtpQRM.
	
	The presence of EPs is further supported by the self-orthogonality of the eigenvectors of the non-Hermitian Hamiltonian~\cite{moiseyev_non-hermitian_2011}. Under an alternative normalization, $\langle L' \vert L' \rangle = \langle R' \vert R' \rangle = 1$, the c-product $\langle L' \vert R' \rangle$ quantifies the overlap between the bra and ket in the biorthogonal basis. As shown in Fig.~\ref{fidelitysus_btp_fig}, this overlap vanishes at EPs, directly reflecting the self-orthogonal nature of the eigenstates. The vanishing c-product thus provides a complementary perspective on the divergent behavior of the fidelity susceptibility near EPs.
	
	\subsection{Dissipative Two-Photon Quantum Rabi Model}
	
	\textsl{Doubly Degenerate Points:} In the dtpQRM, the conservation of parity $\Pi$ allows for the emergence of two distinct types of level intersections: doubly degenerate points—well-established in the Hermitian tpQRM—and EPs, which are characteristic features of non-Hermitian systems.
	
	\begin{figure}[tbp]
		\centering
		\includegraphics[width=1.0\linewidth]{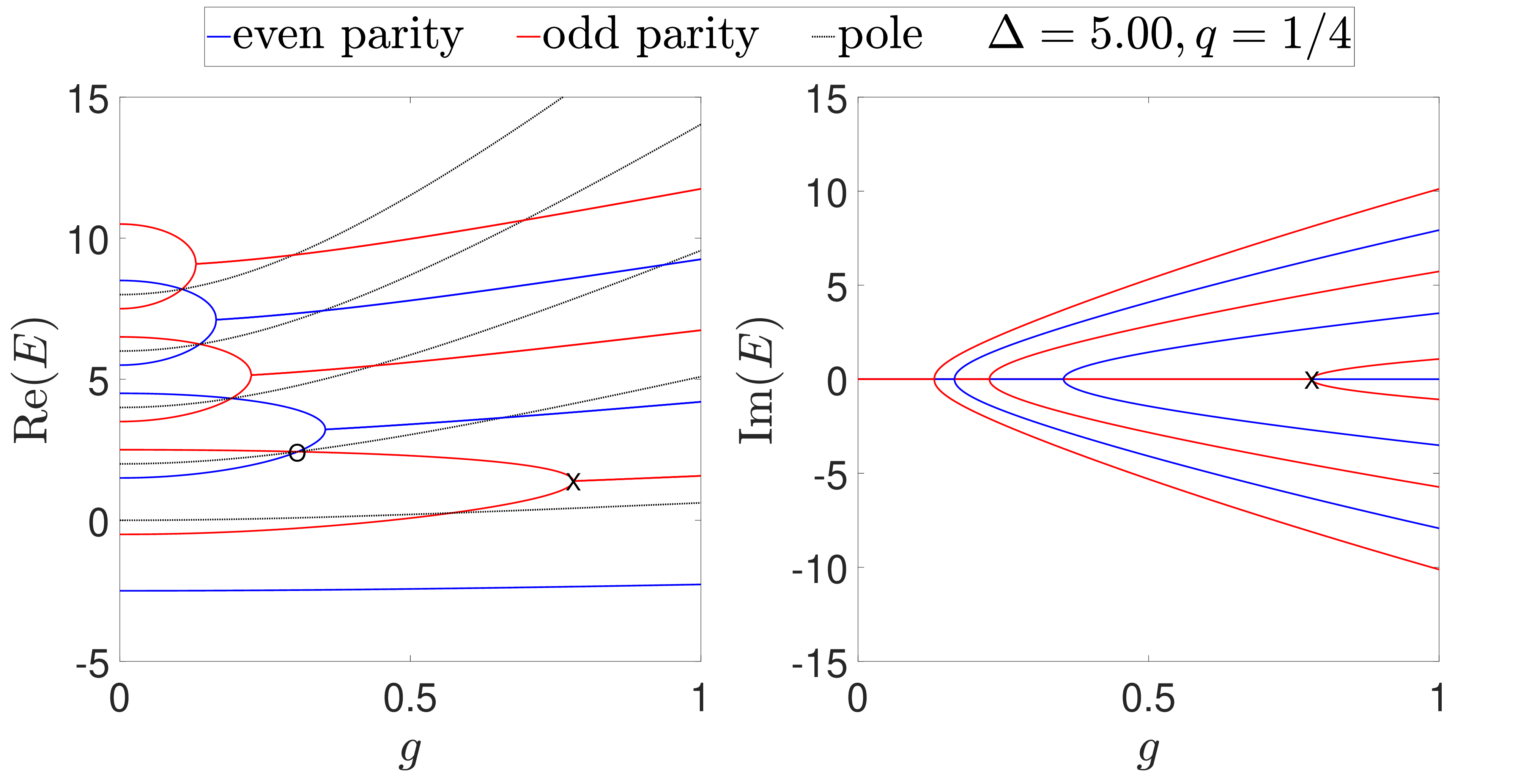}
		\caption{Real (left) and imaginary (right) parts of the lowest few eigenvalues of the dtpQRM as functions of the coupling strength $g$. Blue (red) lines indicate even (odd) $\Pi$-parity levels obtained via exact diagonalization. Dashed lines denote the pole positions $E_{n}^{(q,\mathrm{pole})}$. Here, $\Delta = 5.00$ and $q = 1/4$.}
		\label{spectrum_dtp_d5.00_fig}
	\end{figure}
	
	The doubly degenerate points in the spectrum can be interpreted as Juddian solutions, where the associated eigenfunctions are of finite dimension~\cite{emary_bogoliubov_2002}. Identifying these points requires a careful analysis of the pole structure, as the parity operator $\Pi$ remains a good quantum number whenever the corresponding eigenstate is non-degenerate.
	
	The pole structure observed in Fig.~\ref{G_dtpQRM_fig} is given by
	\begin{equation}
		E_{n}^{(q,\mathrm{pole})} = 2(n + q)\gamma - \frac{1}{2},
	\end{equation}
	where the $G$-function diverges. The first pole appears at $E = 2q\gamma - \frac{1}{2}$, and subsequent poles are evenly spaced by $2\gamma$. Unlike the Hermitian case, where spectral collapse occurs at $g_c = 1/2$ due to vanishing pole spacing~\cite{duan_two-photon_2016}, the dtpQRM does not exhibit such collapse. This is because the pole spacing remains finite for all $g$, as $\gamma = \sqrt{1 + 4g^2}$ is strictly positive. As shown in Fig.~\ref{spectrum_dtp_d5.00_fig}, discrete eigenvalues persist even in the strong-coupling regime, in sharp contrast to the Hermitian tpQRM where the spectrum collapses into a continuum.
	
	At the $n$th pole, where $E = E_{n}^{(q,\mathrm{pole})}$, $e_n^{(q)}$~\eqref{recur_dtpQRM_en} is well-defined only if the following condition is satisfied:
	\begin{eqnarray}
		f_n^{(q)} &=& \frac{ \left[ 2(n - 1 + q)(2 - \gamma^2) - \gamma \left( \frac{1}{2} + E \right) \right] f_{n-1}^{(q)} }{ 8g \left( n + q - \frac{1}{4} \right) \left( n + q - \frac{3}{4} \right) } \nonumber \\
		&& + \frac{ 2g f_{n-2}^{(q)} - \frac{\Delta}{2} \gamma e_{n-1}^{(q)} }{ 8g \left( n + q - \frac{1}{4} \right) \left( n + q - \frac{3}{4} \right) } = 0.
	\end{eqnarray}
	By iterating the recurrence relations \eqref{recur_dtpQRM}, one can determine the corresponding coupling strength $g_n^{(q)}$ at which a doubly degenerate state appears on the $n$th pole line. At this point, both the numerator and the denominator of $e_n^{(q)}$ vanish simultaneously, rendering $e_n^{(q)}$ arbitrary. If we choose
	\begin{equation}
		e_n^{(q)} = \frac{4g f_{n-1}^{(q)}}{\Delta \gamma},
	\end{equation}
	then it follows from the recurrence relation that $f_{n+1}^{(q)} = 0$, and consequently all higher-order coefficients vanish: $f_{m \geqslant n}^{(q)} = 0$ and $e_{m > n}^{(q)} = 0$. This truncation implies that the corresponding eigenfunction contains only a finite number of terms, signifying a Juddian solution~\cite{emary_bogoliubov_2002}.
	
	Since $f_0^{(q)}$ is set to $1$, the first doubly degenerate point appears on the first pole line. For the case of $q = 1/4$, the degeneracy point corresponding to the crossing of the second and third excited states in the eigenvalue spectrum is located at
	\begin{eqnarray}
		g_{1}^{(1/4)} &=& \frac{1}{4} \sqrt{ \frac{ \Delta^2 - 16 }{6} }, \nonumber \\
		E_{1}^{(1/4)} &=& \frac{5}{2} \sqrt{ \frac{ \Delta^2 + 8 }{24} } - \frac{1}{2}.
	\end{eqnarray}
	Such a doubly degenerate solution exists only when $\Delta > 4$. At this point, adjacent energy levels with different $\Pi$ parities intersect precisely on the pole line, with the second and third levels crossing at $g = g_{1}^{(1/4)}$ and $E = E_{1}^{(1/4)}$, as indicated by the circle in Fig.~\ref{spectrum_dtp_d5.00_fig}. Remarkably, the location of this Juddian degeneracy coincides exactly with that in the Hermitian tpQRM.
	
	Therefore, doubly degenerate eigenstates can emerge within the $\mathcal{PT}$-symmetric phase under specific parametric conditions. Moreover, Fig.~\ref{spectrum_dtp_d5.00_fig} also reveals the presence of EPs, where two real eigenvalues coalesce. In contrast to the btpQRM, each eigenvalue branch in the dtpQRM hosts at most one EP, and once a level enters the $\mathcal{PT}$-broken phase, it does not return to the $\mathcal{PT}$-symmetric regime as the coupling strength $g$ increases.
	
	\begin{figure}[tbp]
		\centering
		\includegraphics[width=1.0\linewidth]{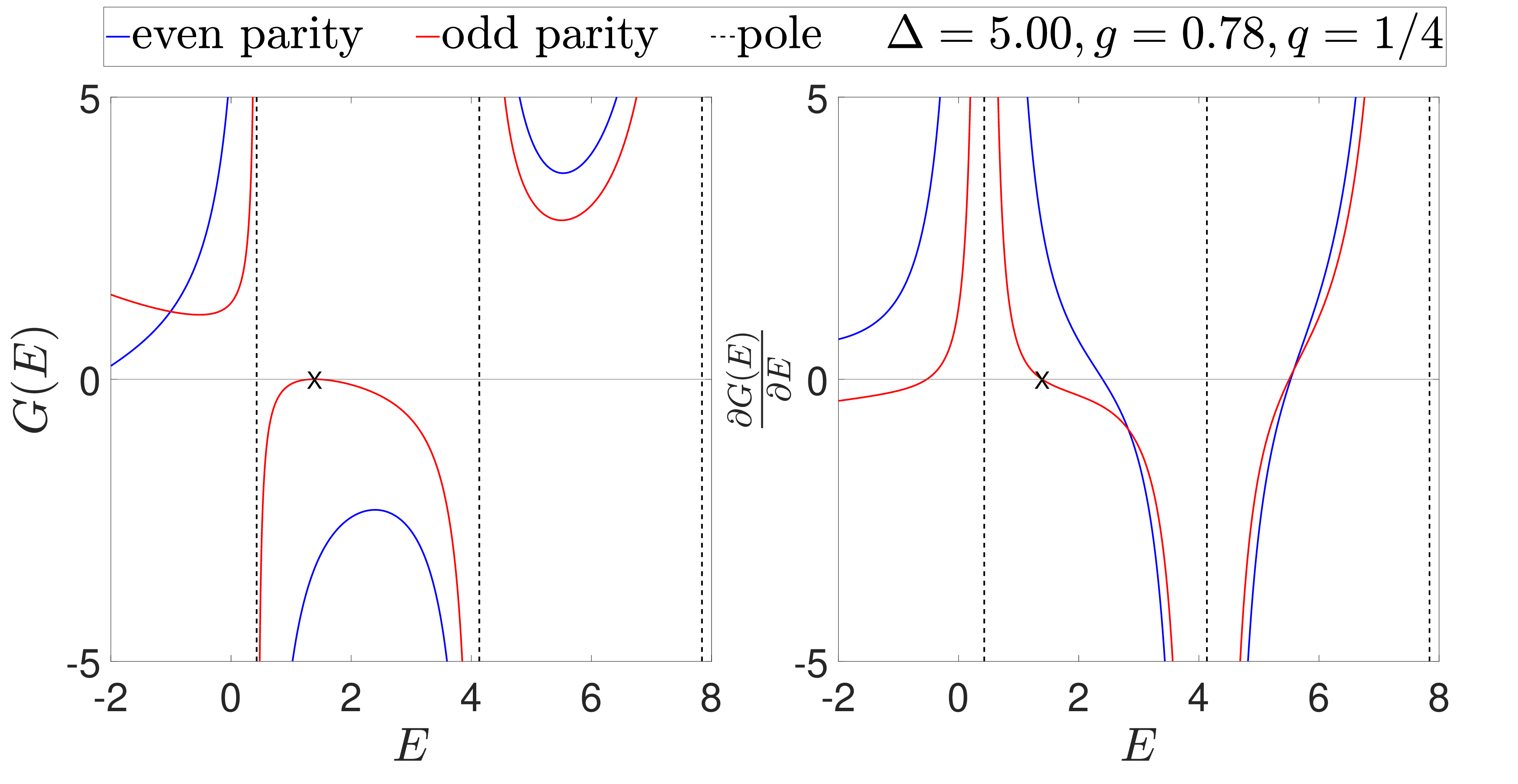}
		\caption{$G$-function curves (left) and their first-order derivatives with respect to $E$ (right) in the real energy regime for the dtpQRM. Parameters: $\Delta = 5.00$, $g = 0.7802$, and $q = 1/4$, corresponding to the EP marked by an “X” in Fig.~\ref{spectrum_dtp_d5.00_fig}. Blue (red) lines represent even (odd) $\Pi$-parity results. Black dashed lines indicate $E_n^{(q,\mathrm{pole})}$.}
		\label{G_EP_dtp_fig}
	\end{figure}
	
	\textsl{Exceptional Points:} EPs occur when two complex-conjugate eigenvalues coalesce, while the parity $\Pi$ of the tpQRM remains conserved. As a result, the $G$-function exhibits a single zero located between two adjacent poles at the EP. As shown in Fig.~\ref{G_EP_dtp_fig}, an EP is characterized by the simultaneous vanishing of both the $G$-function and its first derivative with respect to $E$. At this point, the two eigenvalues merge into a single value, determined by the zero of the $G$-function, and the corresponding eigenfunctions $\vert \Psi_\mathrm{dtp}^{(q)} \rangle$ coalesce. These coalesced states share the same expansion coefficients $e_n^{(q)}$ and $f_n^{(q)}$, and possess the same $\Pi$ parity. Importantly, because EPs do not lie on the pole lines, their associated eigenfunctions cannot be expressed in a finite-dimensional basis—unlike Juddian-type solutions. Furthermore, in contrast to the btpQRM, $\mathcal{PT}$ symmetry cannot be restored once broken, and the imaginary part of the eigenvalue may continue to grow without bound as $g$ increases.
	
	In contrast, doubly degenerate points, which arise due to the conserved parity $\Pi$, occur precisely on the pole lines. At these points, two distinct eigenstates with opposite $\Pi$ parities coexist. However, such degeneracies do not correspond to zeros of the $G$-function, as the $G$-function is parity-resolved and thus cannot capture level crossings between states of differing parity.
	
	\begin{figure}[tbp]
		\centering
		\includegraphics[width=1.0\linewidth]{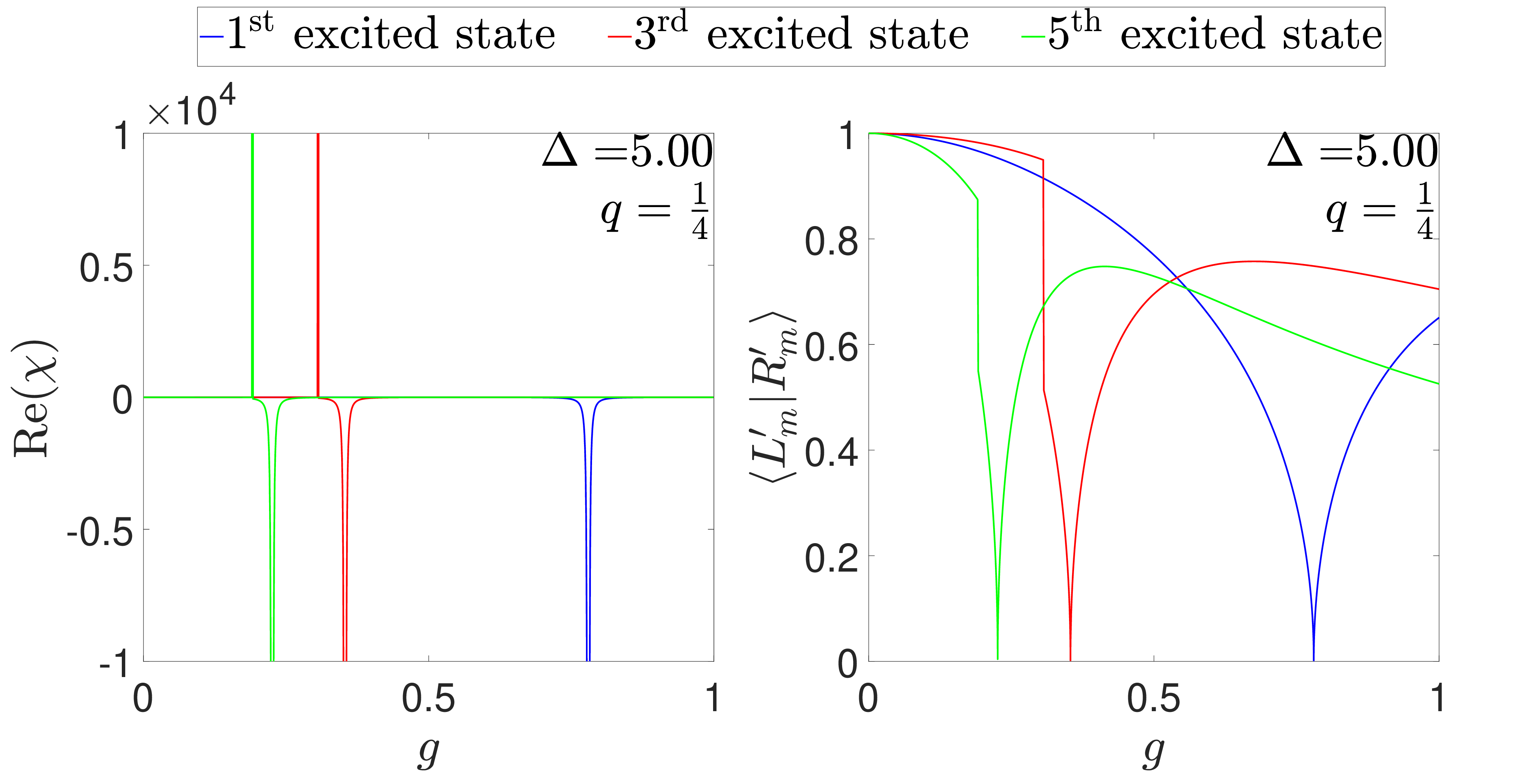}
		\caption{Identification of the two types of level intersections in the dtpQRM under the same parameters as in Fig.~\ref{spectrum_dtp_d5.00_fig}. The real part of the fidelity susceptibility (left) and the c-product $\langle L' \vert R' \rangle$ (right) are plotted as functions of the coupling strength $g$. Blue, red, and green lines correspond to the $1^{\mathrm{st}}$, $3^{\mathrm{rd}}$, and $5^{\mathrm{th}}$ excited states, respectively, with $q = \frac{1}{4}$ and $\Delta = 5.00$.}
		\label{fidelitysus_dtp_fig}
	\end{figure}
	
	As shown in Fig.~\ref{fidelitysus_dtp_fig}, the real part of the fidelity susceptibility, $\mathrm{Re}(\chi)$, provides a clear distinction between EPs and doubly degenerate points. Specifically, $\mathrm{Re}(\chi)$ exhibits a pronounced negative peak in the vicinity of EPs, while showing a significant positive peak near doubly degenerate points.
	
	This distinction stems from the differing behaviors of eigenfunctions in the two scenarios. Adjacent eigenvalue levels with the same $\Pi$ parity evolve continuously as $g$ increases, except at EPs, where they coalesce. In such cases, the fidelity $\mathcal{F}$ can exceed the conventional range $[0,1]$ due to the conditions $\langle L \vert L \rangle = \langle R \vert R \rangle$, which diverge near EPs as a result of self-orthogonality. Consequently, values of $|\mathcal{F}| > 1$ frequently occur in the vicinity of EPs, leading to a negative divergence of $\mathrm{Re}(\chi)$~\cite{tzeng_hunting_2021}.
	
	In contrast, at a doubly degenerate point, the eigenfunction undergoes a parity switching between two states with opposite $\Pi$ parities. This parity switching causes the overlap $\mathcal{F}$ to vanish exactly at the degenerate point, resulting in a positive divergence of $\mathrm{Re}(\chi)$. Notably, the peak in $\mathrm{Re}(\chi)$ at a doubly degenerate point is significantly sharper and narrower than that at an EP, as the divergence originates from symmetry switching rather than eigenstate coalescence. In practice, pinpointing the exact locations of EPs and doubly degenerate points remains challenging. Nevertheless, the broad and pronounced peak structures associated with EPs may provide enhanced experimental accessibility, particularly for applications such as quantum sensing.
	
	The presence of EPs is further substantiated by the self-orthogonality of eigenvectors in the non-Hermitian Hamiltonian~\cite{moiseyev_non-hermitian_2011}. As illustrated in Fig.~\ref{fidelitysus_dtp_fig}, the c-product $\langle L' \vert R' \rangle$ vanishes precisely at EPs, highlighting the self-orthogonal nature of the coalesced eigenstates. Moreover, it exhibits abrupt discontinuities at doubly degenerate points, corresponding to the parity-switching behavior of the eigenfunctions at those intersections. These observations confirm that the c-product serves as a reliable diagnostic tool for distinguishing EPs from doubly degenerate points.
	
	\section{Dynamics of the Two Models}
	
	In this section, we compare the dynamical behaviors of the btpQRM and dtpQRM. Specifically, we compute the time evolution of the qubit population $\langle W \rangle$ and the average photon number $\langle n \rangle = \langle a^\dagger a \rangle$, starting from an initial state in which the qubit is excited and the field is in the vacuum state. The photonic Fock space is truncated at a maximum photon number of 200. Since the Hamiltonian~\eqref{H_dtp} involves a rotation by $\pi/4$ around the $y$ axis in spin space, the qubit population is expressed as
	\begin{equation}
		\langle W \rangle = \frac{1 - \langle \sigma_{x} \rangle}{2}.
	\end{equation}
	
	\subsection{Biased Two-Photon Quantum Rabi Model} \label{sec_dyna_btp}
	
	\begin{figure}[tbp]
		\centering
		\subfloat[\label{dynamic_btp_g010_fig}]{\includegraphics[width=1.0\linewidth]{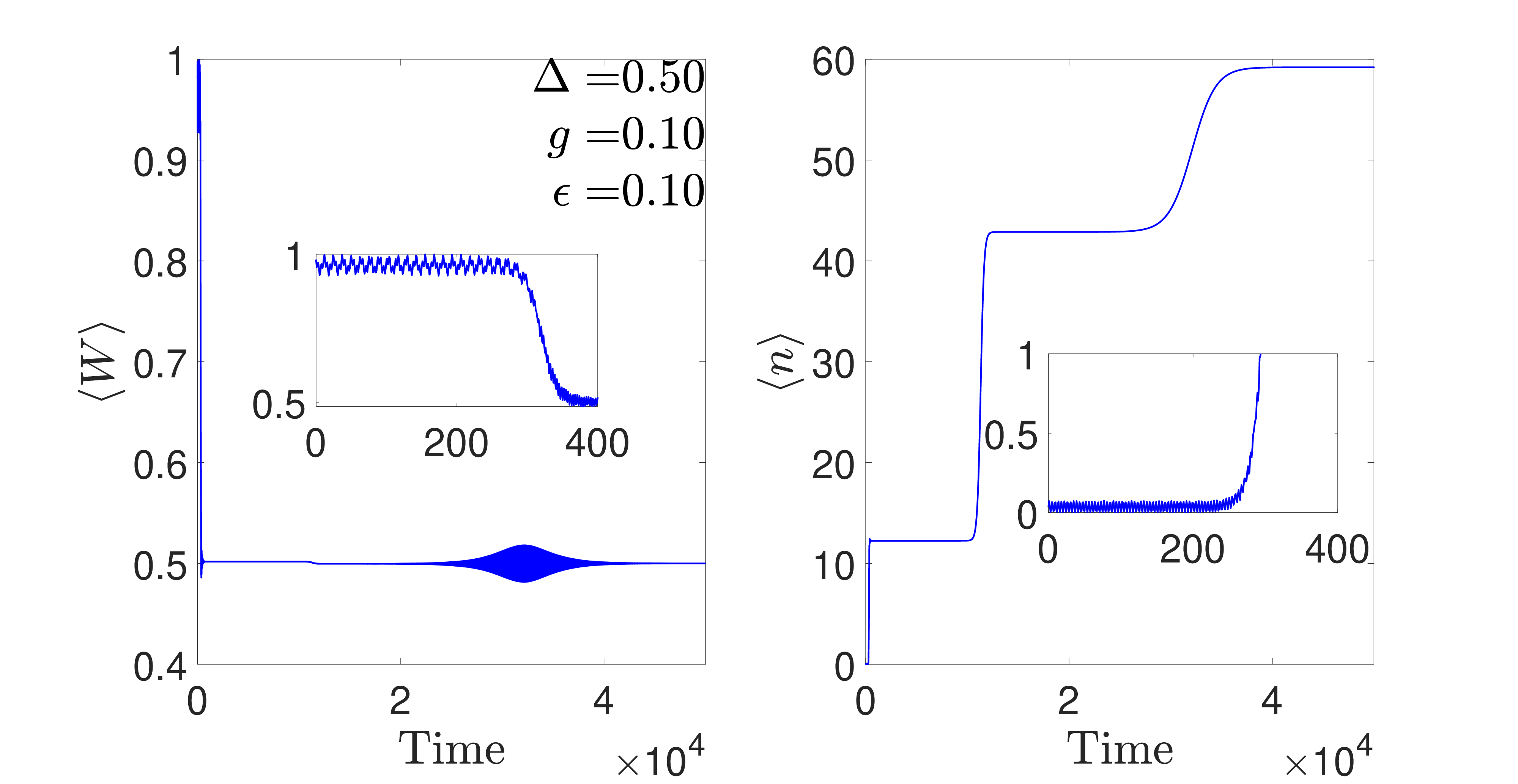}}
		\newline
		\subfloat[\label{dynamic_btp_g025_fig}]{\includegraphics[width=1.0\linewidth]{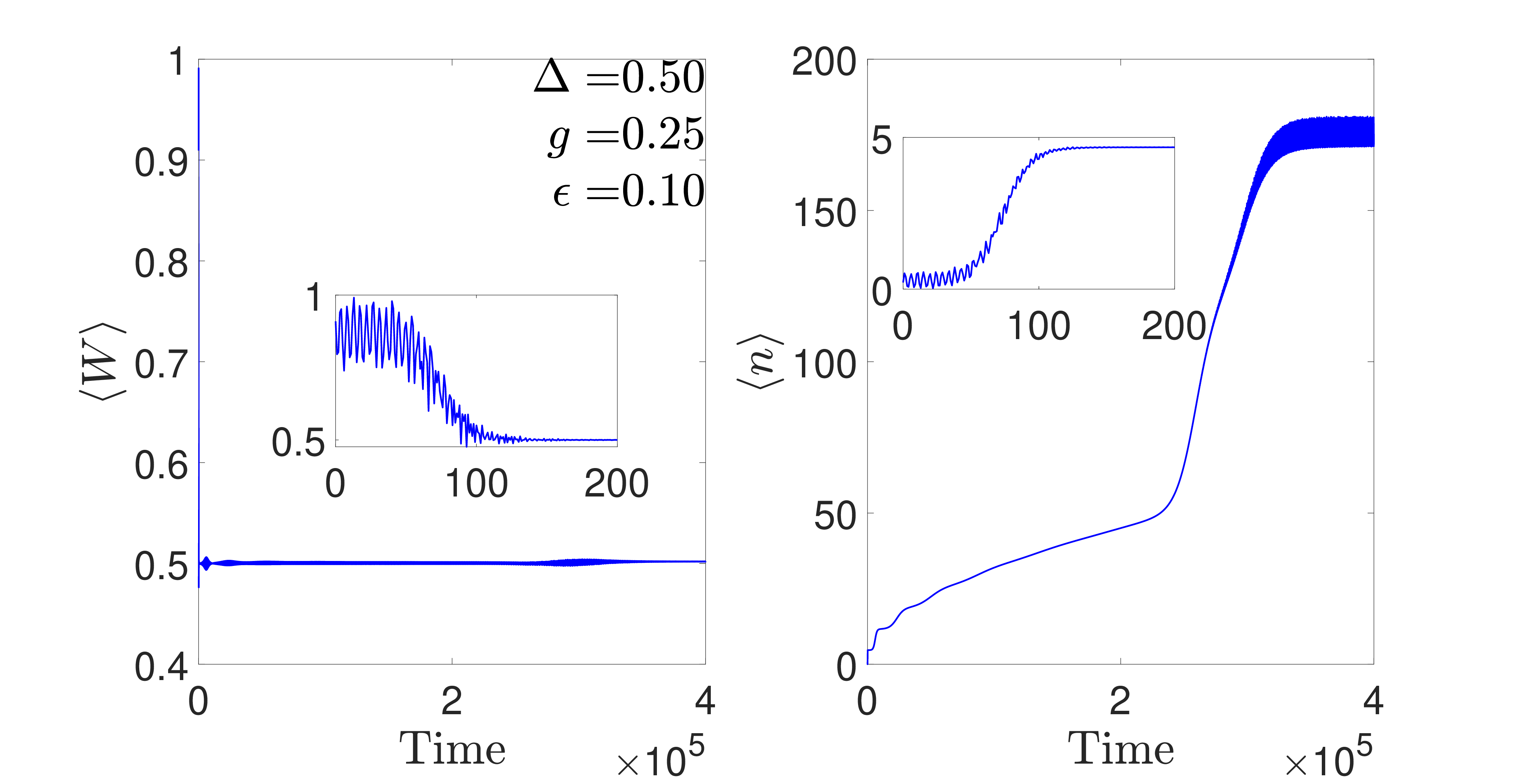}}
		\caption{Time evolution of the qubit population $\langle W \rangle$ (left) and the average photon number $\langle n \rangle$ (right) in the btpQRM. Parameters are $\Delta = 0.50$ and $\epsilon = 0.10$, with $g = 0.10$ in (a) and $g = 0.25$ in (b), corresponding to parameters mearked by the green lines in Fig.~\ref{sepctra_bias_delta0.50_h_fig}. Insets provide magnified views for improved clarity.}
		\label{dynamic_btp_fig}
	\end{figure}
	
	We begin by examining the dynamical evolution of the qubit population and the average photon number at $g = 0.10$ (see Fig.~\ref{dynamic_btp_g010_fig}). As indicated by the first green line in Fig.~\ref{sepctra_bias_delta0.50_h_fig}, the lowest few eigenstates reside within the $\mathcal{PT}$-symmetric phase, whereas $\mathcal{PT}$-symmetry breaking occurs only among higher excited states. In the short-time regime, both observables exhibit regular oscillations around their initial values, closely resembling the behavior observed in Hermitian systems. However, at longer times, the average photon number displays a step-like increase and eventually saturates to a steady value.
	
	This feature is a hallmark of non-Hermitian dynamics, arising from the interplay of gain, loss, and the non-conservation of probability amplitudes intrinsic to the system. Within the $\mathcal{PT}$-broken regime, the presence of complex eigenvalues implies that the corresponding eigenfunctions are no longer orthogonal. Consequently, a low-lying excited state can dynamically evolve into a higher-lying state that also resides in the $\mathcal{PT}$-broken regime. The imaginary components of the eigenvalues represent effective gain and loss: due to their complex-conjugate structure, one eigenmode undergoes exponential decay, while the other grows exponentially in time.
	
	As shown in Fig.~\ref{sepctra_bias_delta0.50_fig}, the $\mathcal{PT}$-broken region forms a narrow arc, indicating a relatively large spacing between adjacent $\mathcal{PT}$-broken levels and small imaginary components—typically less than $\epsilon/2$. This spectral structure gives rise to the stepwise increase observed in the average photon number. Initially, the system evolves into the lowest-lying $\mathcal{PT}$-broken state, whose amplitude gradually dominates the dynamics, resulting in the formation of the first plateau. As time progresses, the system transitions to the next higher $\mathcal{PT}$-broken state, leading to a second plateau, and this process continues sequentially until the state with the largest imaginary part within the truncated Hilbert space becomes dominant. At this point, the system reaches a dynamical steady state. In the absence of truncation, however, the dynamics would ultimately diverge toward a state with an infinite number of photons.
	
	When $g = 0.25$ is selected, the second green line in Fig.~\ref{sepctra_bias_delta0.50_h_fig} shows that the fourth and fifth excited states reside within the $\mathcal{PT}$-broken regime. As depicted in Fig.~\ref{dynamic_btp_g025_fig}, the initial oscillatory behavior around the prepared state rapidly decays, and a pronounced first plateau emerges in the average photon number around the value of 5. This matches the average photon occupations of the fourth and fifth eigenstates, confirming our theoretical expectation. Furthermore, at longer times, the average photon number continues to increase and eventually exhibits oscillatory behavior near the truncation boundary. This suggests that additional $\mathcal{PT}$-broken states exist near the edge of the truncated Hilbert space. Several of these high-lying broken states possess imaginary parts close to $\epsilon/2$, resulting in nearly same amplification and prominent oscillations in the high-photon-number regime.
	
	\subsection{Dissipative Two-Photon Quantum Rabi Model}
	
	\begin{figure}[tbp]
		\centering
		\subfloat[\label{dynamic_dtp_g005_fig}]{\includegraphics[width=1.0\linewidth]{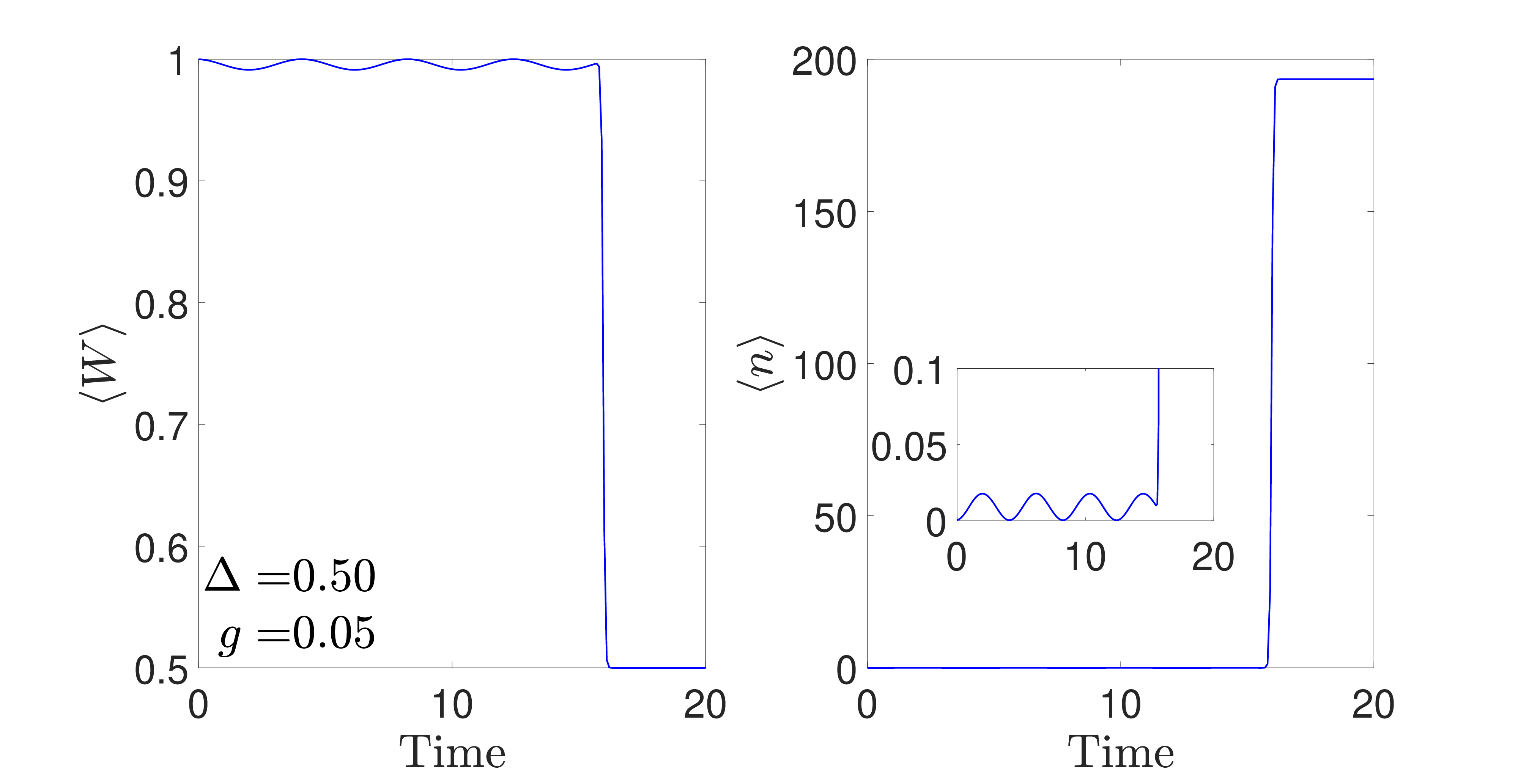}}
		\newline
		\subfloat[\label{dynamic_dtp_g025_fig}]{\includegraphics[width=1.0\linewidth]{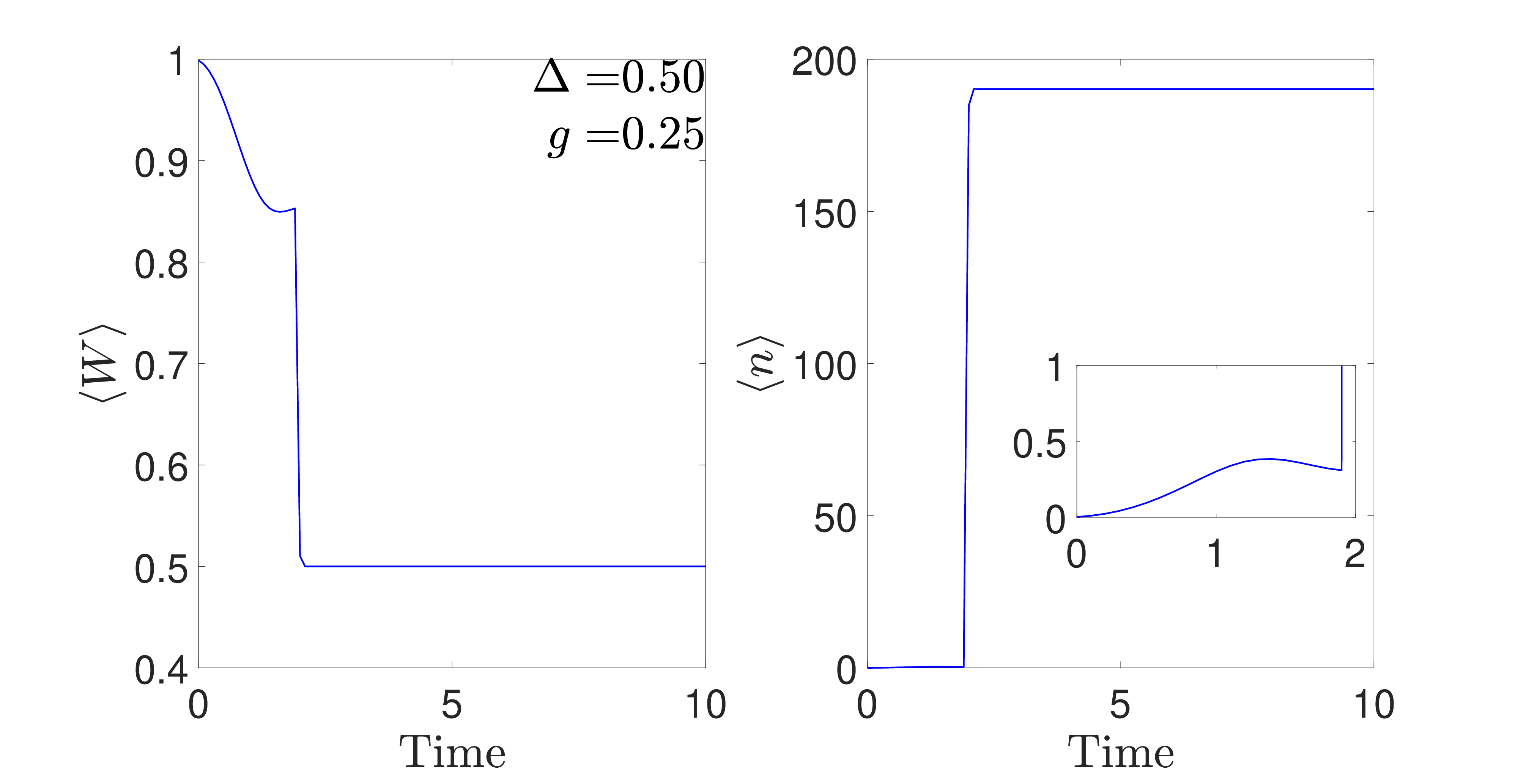}}
		\caption{Time evolution of the qubit population $\langle W \rangle$ (left) and the average photon number $\langle n \rangle$ (right) in the dtpQRM, with $\Delta = 0.50$, $g = 0.05$ in (a) and $g = 0.25$ in (b). Insets display magnified views of the initial dynamics.}
		\label{dynamic_dtp_fig}
	\end{figure}
	
	In stark contrast to the btpQRM, the dtpQRM exhibits a qualitatively distinct dynamical behavior. As shown in Fig.~\ref{dynamic_dtp_fig}, the initial oscillations in both the qubit population and the average photon number decay rapidly, vanishing within a short timescale. Notably, the characteristic step-like structure observed in the btpQRM is entirely absent. Instead, the system undergoes a sharp, single-stage relaxation into a dynamical steady state. This markedly different behavior can be attributed to the underlying spectral properties of the dtpQRM. In particular, each eigenvalue level supports at most one EP, beyond which the system transitions irreversibly into the $\mathcal{PT}$-broken phase. Once this transition occurs, the eigenstate does not revert to the $\mathcal{PT}$-symmetric regime, even as the coupling strength $g$ continues to increase. Moreover, as one moves to higher excited states, the corresponding EPs systematically shift toward lower values of $g$, asymptotically approaching $g = 0$. As a result, for any nonzero $g$, a significant portion of the higher-energy spectrum lies within the $\mathcal{PT}$-broken regime.
	
	This implies that, for any fixed value of $g$, there exists a threshold excitation level beyond which all higher-energy eigenstates reside within the $\mathcal{PT}$-broken phase. Consequently, a wavefunction initially localized in a low-lying $\mathcal{PT}$-broken state can continuously evolve into successively higher-lying $\mathcal{PT}$-broken states. Since the imaginary components of the eigenvalues grow with increasing excitation level, the system dynamically converges toward the eigenstate associated with the largest imaginary part—namely, the highest level supported by the truncated Hilbert space. As a result, the final average photon number asymptotically approaches the truncation boundary. This mechanism also explains the observed dependence of the dynamical evolution on both the coupling strength $g$ and the photon-number cutoff: increasing either parameter enhances the rate at which the system settles into its steady state, as more highly unstable $\mathcal{PT}$-broken states become dynamically accessible
	
	\section{CONCLUSION} 
	
	In this work, we have systematically studied two $\mathcal{PT}$-symmetric non-Hermitian variants of the tpQRM: the btpQRM and the dtpQRM. These models exhibit distinct physical mechanisms—balanced gain and loss via imaginary qubit bias in the btpQRM, and engineered dissipation through imaginary coupling in the dtpQRM—while preserving the essential $\mathcal{PT}$ symmetry.
	
	For both models, we constructed exact solutions using a combination of Bogoliubov transformations and $\text{su}(1,1)$ algebra. This allowed us to construct $G$-functions whose zeros determine the eigenvalue spectra. In the btpQRM, we found that spectral collapse persists at the critical coupling strength $g_c = 1/2$, even in the non-Hermitian setting. The introduction of a small imaginary bias leads to the appearance of EPs distributed along the pole lines, which correspond to the Juddian-type doubly degenerate points in the Hermitian model. This correspondence was further supported through an adiabatic approximation, which accurately captured the $\mathcal{PT}$-broken thresholds and their dependence on the qubit bias $\epsilon$.
	
	In the dtpQRM, spectral collapse is absent due to the persistence of finite pole spacing in the $G$-function. Nevertheless, we identified two distinct types of level intersections: EPs between states of the same parity, and Juddian-type doubly degenerate points between states of opposite parity. We analytically located the Juddian solutions via recursion relations and showed that EPs arise off the pole lines, where both the $G$-function and its derivative vanish. The fidelity susceptibility and c-product provided powerful tools for distinguishing between EPs (characterized by self-orthogonality and negative fidelity divergence) and doubly degenerate points (marked by symmetry switching and positive fidelity divergence).
	
	The dynamical behavior of the two models further highlights their qualitative differences. In the btpQRM, the stepwise evolution and long-time saturation of observables reflect the discrete nature of $\mathcal{PT}$-broken modes and the relatively isolated EPs. In contrast, the dtpQRM dynamics exhibit rapid convergence to steady states due to the unbounded growth of imaginary parts in the spectrum and the absence of symmetry restoration once a level becomes broken.
	
	In summary, our study demonstrates that non-Hermitian extensions of the tpQRM yield rich and contrasting spectral and dynamical phenomena depending on the manner of $\mathcal{PT}$-symmetry implementation. The interplay between EPs, Juddian solutions, and dynamical transitions offers fertile ground for both theoretical exploration and experimental realization in platforms such as circuit QED, trapped ions, and cold atomic systems. Our exact analytical framework provides concrete guidance for identifying spectral features and dynamical signatures associated with non-Hermitian quantum criticality and $\mathcal{PT}$-symmetry breaking. These insights deepen our understanding of non-Hermitian QRMs and pave the way for experimental studies in cavity QED, trapped ions, and related systems.
	
	\acknowledgments
	This work is supported in part by the National Key R$\&$D  Program of China (Grant No. 2024YFA1408900).
	
	%\bibliographystyle{apsrev4-2}
	%\bibliography{refs}

\begin{thebibliography}{68}%
		\makeatletter
		\providecommand \@ifxundefined [1]{%
			\@ifx{#1\undefined}
		}%
		\providecommand \@ifnum [1]{%
			\ifnum #1\expandafter \@firstoftwo
			\else \expandafter \@secondoftwo
			\fi
		}%
		\providecommand \@ifx [1]{%
			\ifx #1\expandafter \@firstoftwo
			\else \expandafter \@secondoftwo
			\fi
		}%
		\providecommand \natexlab [1]{#1}%
		\providecommand \enquote  [1]{``#1''}%
		\providecommand \bibnamefont  [1]{#1}%
		\providecommand \bibfnamefont [1]{#1}%
		\providecommand \citenamefont [1]{#1}%
		\providecommand \href@noop [0]{\@secondoftwo}%
		\providecommand \href [0]{\begingroup \@sanitize@url \@href}%
		\providecommand \@href[1]{\@@startlink{#1}\@@href}%
		\providecommand \@@href[1]{\endgroup#1\@@endlink}%
		\providecommand \@sanitize@url [0]{\catcode `\\12\catcode `\$12\catcode
			`\&12\catcode `\#12\catcode `\^12\catcode `\_12\catcode `\%12\relax}%
		\providecommand \@@startlink[1]{}%
		\providecommand \@@endlink[0]{}%
		\providecommand \url  [0]{\begingroup\@sanitize@url \@url }%
		\providecommand \@url [1]{\endgroup\@href {#1}{\urlprefix }}%
		\providecommand \urlprefix  [0]{URL }%
		\providecommand \Eprint [0]{\href }%
		\providecommand \doibase [0]{https://doi.org/}%
		\providecommand \selectlanguage [0]{\@gobble}%
		\providecommand \bibinfo  [0]{\@secondoftwo}%
		\providecommand \bibfield  [0]{\@secondoftwo}%
		\providecommand \translation [1]{[#1]}%
		\providecommand \BibitemOpen [0]{}%
		\providecommand \bibitemStop [0]{}%
		\providecommand \bibitemNoStop [0]{.\EOS\space}%
		\providecommand \EOS [0]{\spacefactor3000\relax}%
		\providecommand \BibitemShut  [1]{\csname bibitem#1\endcsname}%
		\let\auto@bib@innerbib\@empty
		%</preamble>
		\bibitem [{\citenamefont {Bender}(2007)}]{bender_making_2007}%
		\BibitemOpen
		\bibfield  {author} {\bibinfo {author} {\bibfnamefont {C.~M.}\ \bibnamefont
				{Bender}},\ }\href {https://doi.org/10.1088/0034-4885/70/6/R03} {\bibfield
			{journal} {\bibinfo  {journal} {Rep. Prog. Phys.}\ }\textbf {\bibinfo
				{volume} {70}},\ \bibinfo {pages} {947} (\bibinfo {year} {2007})}\BibitemShut
		{NoStop}%
		\bibitem [{\citenamefont {Moiseyev}(2011)}]{moiseyev_non-hermitian_2011}%
		\BibitemOpen
		\bibfield  {author} {\bibinfo {author} {\bibfnamefont {N.}~\bibnamefont
				{Moiseyev}},\ }\href {https://doi.org/10.1017/CBO9780511976186} {\emph
			{\bibinfo {title} {Non-Hermitian Quantum Mechanics}}}\ (\bibinfo  {publisher}
		{Cambridge University Press},\ \bibinfo {year} {2011})\BibitemShut {NoStop}%
		\bibitem [{\citenamefont {Ashida}\ \emph {et~al.}(2020)\citenamefont {Ashida},
			\citenamefont {Gong},\ and\ \citenamefont
			{Ueda}}]{ashida_non-hermitian_2020}%
		\BibitemOpen
		\bibfield  {author} {\bibinfo {author} {\bibfnamefont {Y.}~\bibnamefont
				{Ashida}}, \bibinfo {author} {\bibfnamefont {Z.}~\bibnamefont {Gong}},\ and\
			\bibinfo {author} {\bibfnamefont {M.}~\bibnamefont {Ueda}},\ }\href
		{https://doi.org/10.1080/00018732.2021.1876991} {\bibfield  {journal}
			{\bibinfo  {journal} {Advances in Physics}\ }\textbf {\bibinfo {volume}
				{69}},\ \bibinfo {pages} {249} (\bibinfo {year} {2020})}\BibitemShut
		{NoStop}%
		\bibitem [{\citenamefont {Dum}\ \emph {et~al.}(1992)\citenamefont {Dum},
			\citenamefont {Zoller},\ and\ \citenamefont {Ritsch}}]{dum_monte_1992}%
		\BibitemOpen
		\bibfield  {author} {\bibinfo {author} {\bibfnamefont {R.}~\bibnamefont
				{Dum}}, \bibinfo {author} {\bibfnamefont {P.}~\bibnamefont {Zoller}},\ and\
			\bibinfo {author} {\bibfnamefont {H.}~\bibnamefont {Ritsch}},\ }\href
		{https://doi.org/10.1103/PhysRevA.45.4879} {\bibfield  {journal} {\bibinfo
				{journal} {Phys. Rev. A}\ }\textbf {\bibinfo {volume} {45}},\ \bibinfo
			{pages} {4879} (\bibinfo {year} {1992})}\BibitemShut {NoStop}%
		\bibitem [{\citenamefont {Otterbach}\ and\ \citenamefont
			{Lemeshko}(2014)}]{otterbach_dissipative_2014}%
		\BibitemOpen
		\bibfield  {author} {\bibinfo {author} {\bibfnamefont {J.}~\bibnamefont
				{Otterbach}}\ and\ \bibinfo {author} {\bibfnamefont {M.}~\bibnamefont
				{Lemeshko}},\ }\href {https://doi.org/10.1103/PhysRevLett.113.070401}
		{\bibfield  {journal} {\bibinfo  {journal} {Phys. Rev. Lett.}\ }\textbf
			{\bibinfo {volume} {113}},\ \bibinfo {pages} {070401} (\bibinfo {year}
			{2014})}\BibitemShut {NoStop}%
		\bibitem [{\citenamefont {Lee}\ and\ \citenamefont
			{Chan}(2014)}]{lee_heralded_2014}%
		\BibitemOpen
		\bibfield  {author} {\bibinfo {author} {\bibfnamefont {T.~E.}\ \bibnamefont
				{Lee}}\ and\ \bibinfo {author} {\bibfnamefont {C.-K.}\ \bibnamefont {Chan}},\
		}\href {https://doi.org/10.1103/PhysRevX.4.041001} {\bibfield  {journal}
			{\bibinfo  {journal} {Phys. Rev. X}\ }\textbf {\bibinfo {volume} {4}},\
			\bibinfo {pages} {041001} (\bibinfo {year} {2014})}\BibitemShut {NoStop}%
		\bibitem [{\citenamefont {Li}\ \emph {et~al.}(2019)\citenamefont {Li},
			\citenamefont {Harter}, \citenamefont {Liu}, \citenamefont {De~Melo},
			\citenamefont {Joglekar},\ and\ \citenamefont {Luo}}]{li_observation_2019}%
		\BibitemOpen
		\bibfield  {author} {\bibinfo {author} {\bibfnamefont {J.}~\bibnamefont
				{Li}}, \bibinfo {author} {\bibfnamefont {A.~K.}\ \bibnamefont {Harter}},
			\bibinfo {author} {\bibfnamefont {J.}~\bibnamefont {Liu}}, \bibinfo {author}
			{\bibfnamefont {L.}~\bibnamefont {De~Melo}}, \bibinfo {author} {\bibfnamefont
				{Y.~N.}\ \bibnamefont {Joglekar}},\ and\ \bibinfo {author} {\bibfnamefont
				{L.}~\bibnamefont {Luo}},\ }\href
		{https://doi.org/10.1038/s41467-019-08596-1} {\bibfield  {journal} {\bibinfo
				{journal} {Nat Commun}\ }\textbf {\bibinfo {volume} {10}},\ \bibinfo {pages}
			{855} (\bibinfo {year} {2019})}\BibitemShut {NoStop}%
		\bibitem [{\citenamefont {Li}\ and\ \citenamefont
			{Xu}(2022)}]{li_non-hermitian_2022}%
		\BibitemOpen
		\bibfield  {author} {\bibinfo {author} {\bibfnamefont {K.}~\bibnamefont
				{Li}}\ and\ \bibinfo {author} {\bibfnamefont {Y.}~\bibnamefont {Xu}},\ }\href
		{https://doi.org/10.1103/PhysRevLett.129.093001} {\bibfield  {journal}
			{\bibinfo  {journal} {Phys. Rev. Lett.}\ }\textbf {\bibinfo {volume} {129}},\
			\bibinfo {pages} {093001} (\bibinfo {year} {2022})}\BibitemShut {NoStop}%
		\bibitem [{\citenamefont {Hatano}\ and\ \citenamefont
			{Nelson}(1996)}]{hatano_localization_1996}%
		\BibitemOpen
		\bibfield  {author} {\bibinfo {author} {\bibfnamefont {N.}~\bibnamefont
				{Hatano}}\ and\ \bibinfo {author} {\bibfnamefont {D.~R.}\ \bibnamefont
				{Nelson}},\ }\href {https://doi.org/10.1103/PhysRevLett.77.570} {\bibfield
			{journal} {\bibinfo  {journal} {Phys. Rev. Lett.}\ }\textbf {\bibinfo
				{volume} {77}},\ \bibinfo {pages} {570} (\bibinfo {year} {1996})}\BibitemShut
		{NoStop}%
		\bibitem [{\citenamefont {Refael}\ \emph {et~al.}(2006)\citenamefont {Refael},
			\citenamefont {Hofstetter},\ and\ \citenamefont
			{Nelson}}]{refael_transverse_2006}%
		\BibitemOpen
		\bibfield  {author} {\bibinfo {author} {\bibfnamefont {G.}~\bibnamefont
				{Refael}}, \bibinfo {author} {\bibfnamefont {W.}~\bibnamefont {Hofstetter}},\
			and\ \bibinfo {author} {\bibfnamefont {D.~R.}\ \bibnamefont {Nelson}},\
		}\href {https://doi.org/10.1103/PhysRevB.74.174520} {\bibfield  {journal}
			{\bibinfo  {journal} {Phys. Rev. B}\ }\textbf {\bibinfo {volume} {74}},\
			\bibinfo {pages} {174520} (\bibinfo {year} {2006})}\BibitemShut {NoStop}%
		\bibitem [{\citenamefont {Longhi}(2013)}]{longhi_non-hermitian_2013}%
		\BibitemOpen
		\bibfield  {author} {\bibinfo {author} {\bibfnamefont {S.}~\bibnamefont
				{Longhi}},\ }\href {https://doi.org/10.1103/PhysRevA.88.062112} {\bibfield
			{journal} {\bibinfo  {journal} {Phys. Rev. A}\ }\textbf {\bibinfo {volume}
				{88}},\ \bibinfo {pages} {062112} (\bibinfo {year} {2013})}\BibitemShut
		{NoStop}%
		\bibitem [{\citenamefont {Neria}\ \emph {et~al.}(1991)\citenamefont {Neria},
			\citenamefont {Nitzan}, \citenamefont {Barnett},\ and\ \citenamefont
			{Landman}}]{neria_quantum_1991}%
		\BibitemOpen
		\bibfield  {author} {\bibinfo {author} {\bibfnamefont {E.}~\bibnamefont
				{Neria}}, \bibinfo {author} {\bibfnamefont {A.}~\bibnamefont {Nitzan}},
			\bibinfo {author} {\bibfnamefont {R.~N.}\ \bibnamefont {Barnett}},\ and\
			\bibinfo {author} {\bibfnamefont {U.}~\bibnamefont {Landman}},\ }\bibfield
		{journal} {\bibinfo  {journal} {Phys. Rev. Lett.}\ }\textbf {\bibinfo
			{volume} {67}},\ \href {https://doi.org/10.1103/PhysRevLett.67.1011}
		{10.1103/PhysRevLett.67.1011} (\bibinfo {year} {1991})\BibitemShut {NoStop}%
		\bibitem [{\citenamefont {Granucci}\ \emph {et~al.}(2010)\citenamefont
			{Granucci}, \citenamefont {Persico},\ and\ \citenamefont
			{Zoccante}}]{granucci_including_2024}%
		\BibitemOpen
		\bibfield  {author} {\bibinfo {author} {\bibfnamefont {G.}~\bibnamefont
				{Granucci}}, \bibinfo {author} {\bibfnamefont {M.}~\bibnamefont {Persico}},\
			and\ \bibinfo {author} {\bibfnamefont {A.}~\bibnamefont {Zoccante}},\
		}\bibfield  {journal} {\bibinfo  {journal} {J. Chem. Phys.}\ }\href
		{https://doi.org/10.1063/1.3489004} {10.1063/1.3489004} (\bibinfo {year}
		{2010})\BibitemShut {NoStop}%
		\bibitem [{\citenamefont {Jaeger}\ \emph {et~al.}(2012)\citenamefont {Jaeger},
			\citenamefont {Fischer},\ and\ \citenamefont
			{Prezhdo}}]{jaeger_decoherence-induced_2024}%
		\BibitemOpen
		\bibfield  {author} {\bibinfo {author} {\bibfnamefont {H.~M.}\ \bibnamefont
				{Jaeger}}, \bibinfo {author} {\bibfnamefont {S.}~\bibnamefont {Fischer}},\
			and\ \bibinfo {author} {\bibfnamefont {O.~V.}\ \bibnamefont {Prezhdo}},\
		}\bibfield  {journal} {\bibinfo  {journal} {J. Chem. Phys.}\ }\href
		{https://doi.org/10.1063/1.4757100} {10.1063/1.4757100} (\bibinfo {year}
		{2012})\BibitemShut {NoStop}%
		\bibitem [{\citenamefont {Wiseman}(1996)}]{HM_Wiseman_1996}%
		\BibitemOpen
		\bibfield  {author} {\bibinfo {author} {\bibfnamefont {H.~M.}\ \bibnamefont
				{Wiseman}},\ }\href {https://doi.org/10.1088/1355-5111/8/1/015} {\bibfield
			{journal} {\bibinfo  {journal} {Quantum and Semiclassical Optics: Journal of
					the European Optical Society Part B}\ }\textbf {\bibinfo {volume} {8}},\
			\bibinfo {pages} {205} (\bibinfo {year} {1996})}\BibitemShut {NoStop}%
		\bibitem [{\citenamefont
			{Mostafazadeh}(2002{\natexlab{a}})}]{mostafazadeh_pseudo-hermiticity_2002}%
		\BibitemOpen
		\bibfield  {author} {\bibinfo {author} {\bibfnamefont {A.}~\bibnamefont
				{Mostafazadeh}},\ }\href {https://doi.org/10.1063/1.1418246} {\bibfield
			{journal} {\bibinfo  {journal} {Journal of Mathematical Physics}\ }\textbf
			{\bibinfo {volume} {43}},\ \bibinfo {pages} {205} (\bibinfo {year}
			{2002}{\natexlab{a}})}\BibitemShut {NoStop}%
		\bibitem [{\citenamefont
			{Mostafazadeh}(2002{\natexlab{b}})}]{mostafazadeh_pseudo-hermiticity_2002-1}%
		\BibitemOpen
		\bibfield  {author} {\bibinfo {author} {\bibfnamefont {A.}~\bibnamefont
				{Mostafazadeh}},\ }\href {https://doi.org/10.1063/1.1461427} {\bibfield
			{journal} {\bibinfo  {journal} {Journal of Mathematical Physics}\ }\textbf
			{\bibinfo {volume} {43}},\ \bibinfo {pages} {2814} (\bibinfo {year}
			{2002}{\natexlab{b}})}\BibitemShut {NoStop}%
		\bibitem [{\citenamefont {Heyl}\ \emph {et~al.}(2013)\citenamefont {Heyl},
			\citenamefont {Polkovnikov},\ and\ \citenamefont
			{Kehrein}}]{heyl_dynamical_2013}%
		\BibitemOpen
		\bibfield  {author} {\bibinfo {author} {\bibfnamefont {M.}~\bibnamefont
				{Heyl}}, \bibinfo {author} {\bibfnamefont {A.}~\bibnamefont {Polkovnikov}},\
			and\ \bibinfo {author} {\bibfnamefont {S.}~\bibnamefont {Kehrein}},\ }\href
		{https://doi.org/10.1103/PhysRevLett.110.135704} {\bibfield  {journal}
			{\bibinfo  {journal} {Phys. Rev. Lett.}\ }\textbf {\bibinfo {volume} {110}},\
			\bibinfo {pages} {135704} (\bibinfo {year} {2013})}\BibitemShut {NoStop}%
		\bibitem [{\citenamefont {Fring}\ and\ \citenamefont
			{Frith}(2017)}]{fring_exact_2017}%
		\BibitemOpen
		\bibfield  {author} {\bibinfo {author} {\bibfnamefont {A.}~\bibnamefont
				{Fring}}\ and\ \bibinfo {author} {\bibfnamefont {T.}~\bibnamefont {Frith}},\
		}\href {https://doi.org/10.1103/PhysRevA.95.010102} {\bibfield  {journal}
			{\bibinfo  {journal} {Phys. Rev. A}\ }\textbf {\bibinfo {volume} {95}},\
			\bibinfo {pages} {010102} (\bibinfo {year} {2017})}\BibitemShut {NoStop}%
		\bibitem [{\citenamefont {Curtright}\ and\ \citenamefont
			{Mezincescu}(2007)}]{curtright_biorthogonal_2007}%
		\BibitemOpen
		\bibfield  {author} {\bibinfo {author} {\bibfnamefont {T.}~\bibnamefont
				{Curtright}}\ and\ \bibinfo {author} {\bibfnamefont {L.}~\bibnamefont
				{Mezincescu}},\ }\href {https://doi.org/10.1063/1.2196243} {\bibfield
			{journal} {\bibinfo  {journal} {Journal of Mathematical Physics}\ }\textbf
			{\bibinfo {volume} {48}},\ \bibinfo {pages} {092106} (\bibinfo {year}
			{2007})}\BibitemShut {NoStop}%
		\bibitem [{\citenamefont {Brody}(2014)}]{brody_biorthogonal_2014}%
		\BibitemOpen
		\bibfield  {author} {\bibinfo {author} {\bibfnamefont {D.~C.}\ \bibnamefont
				{Brody}},\ }\href {https://doi.org/10.1088/1751-8113/47/3/035305} {\bibfield
			{journal} {\bibinfo  {journal} {J. Phys. A: Math. Theor.}\ }\textbf {\bibinfo
				{volume} {47}},\ \bibinfo {pages} {035305} (\bibinfo {year}
			{2014})}\BibitemShut {NoStop}%
		\bibitem [{\citenamefont {Tzeng}\ \emph {et~al.}(2021)\citenamefont {Tzeng},
			\citenamefont {Ju}, \citenamefont {Chen},\ and\ \citenamefont
			{Huang}}]{tzeng_hunting_2021}%
		\BibitemOpen
		\bibfield  {author} {\bibinfo {author} {\bibfnamefont {Y.-C.}\ \bibnamefont
				{Tzeng}}, \bibinfo {author} {\bibfnamefont {C.-Y.}\ \bibnamefont {Ju}},
			\bibinfo {author} {\bibfnamefont {G.-Y.}\ \bibnamefont {Chen}},\ and\
			\bibinfo {author} {\bibfnamefont {W.-M.}\ \bibnamefont {Huang}},\ }\href
		{https://doi.org/10.1103/PhysRevResearch.3.013015} {\bibfield  {journal}
			{\bibinfo  {journal} {Phys. Rev. Research}\ }\textbf {\bibinfo {volume}
				{3}},\ \bibinfo {pages} {013015} (\bibinfo {year} {2021})}\BibitemShut
		{NoStop}%
		\bibitem [{\citenamefont {Bender}\ and\ \citenamefont
			{Boettcher}(1998)}]{bender_real_1998}%
		\BibitemOpen
		\bibfield  {author} {\bibinfo {author} {\bibfnamefont {C.~M.}\ \bibnamefont
				{Bender}}\ and\ \bibinfo {author} {\bibfnamefont {S.}~\bibnamefont
				{Boettcher}},\ }\href {https://doi.org/10.1103/PhysRevLett.80.5243}
		{\bibfield  {journal} {\bibinfo  {journal} {Phys. Rev. Lett.}\ }\textbf
			{\bibinfo {volume} {80}},\ \bibinfo {pages} {5243} (\bibinfo {year}
			{1998})}\BibitemShut {NoStop}%
		\bibitem [{\citenamefont {Klaiman}\ \emph {et~al.}(2008)\citenamefont
			{Klaiman}, \citenamefont {Günther},\ and\ \citenamefont
			{Moiseyev}}]{klaiman_visualization_2008}%
		\BibitemOpen
		\bibfield  {author} {\bibinfo {author} {\bibfnamefont {S.}~\bibnamefont
				{Klaiman}}, \bibinfo {author} {\bibfnamefont {U.}~\bibnamefont {Günther}},\
			and\ \bibinfo {author} {\bibfnamefont {N.}~\bibnamefont {Moiseyev}},\ }\href
		{https://doi.org/10.1103/PhysRevLett.101.080402} {\bibfield  {journal}
			{\bibinfo  {journal} {Phys. Rev. Lett.}\ }\textbf {\bibinfo {volume} {101}},\
			\bibinfo {pages} {080402} (\bibinfo {year} {2008})}\BibitemShut {NoStop}%
		\bibitem [{\citenamefont {Rüter}\ \emph {et~al.}(2010)\citenamefont {Rüter},
			\citenamefont {Makris}, \citenamefont {El-Ganainy}, \citenamefont
			{Christodoulides}, \citenamefont {Segev},\ and\ \citenamefont
			{Kip}}]{ruter_observation_2010}%
		\BibitemOpen
		\bibfield  {author} {\bibinfo {author} {\bibfnamefont {C.~E.}\ \bibnamefont
				{Rüter}}, \bibinfo {author} {\bibfnamefont {K.~G.}\ \bibnamefont {Makris}},
			\bibinfo {author} {\bibfnamefont {R.}~\bibnamefont {El-Ganainy}}, \bibinfo
			{author} {\bibfnamefont {D.~N.}\ \bibnamefont {Christodoulides}}, \bibinfo
			{author} {\bibfnamefont {M.}~\bibnamefont {Segev}},\ and\ \bibinfo {author}
			{\bibfnamefont {D.}~\bibnamefont {Kip}},\ }\href
		{https://doi.org/10.1038/nphys1515} {\bibfield  {journal} {\bibinfo
				{journal} {Nature Phys}\ }\textbf {\bibinfo {volume} {6}},\ \bibinfo {pages}
			{192} (\bibinfo {year} {2010})}\BibitemShut {NoStop}%
		\bibitem [{\citenamefont {Bender}(2015)}]{bender_pt_2015}%
		\BibitemOpen
		\bibfield  {author} {\bibinfo {author} {\bibfnamefont {C.~M.}\ \bibnamefont
				{Bender}},\ }\href {https://doi.org/10.1088/1742-6596/631/1/012002}
		{\bibfield  {journal} {\bibinfo  {journal} {J. Phys.: Conf. Ser.}\ }\textbf
			{\bibinfo {volume} {631}},\ \bibinfo {pages} {012002} (\bibinfo {year}
			{2015})}\BibitemShut {NoStop}%
		\bibitem [{\citenamefont {El-Ganainy}\ \emph {et~al.}(2018)\citenamefont
			{El-Ganainy}, \citenamefont {Makris}, \citenamefont {Khajavikhan},
			\citenamefont {Musslimani}, \citenamefont {Rotter},\ and\ \citenamefont
			{Christodoulides}}]{el-ganainy_non-hermitian_2018}%
		\BibitemOpen
		\bibfield  {author} {\bibinfo {author} {\bibfnamefont {R.}~\bibnamefont
				{El-Ganainy}}, \bibinfo {author} {\bibfnamefont {K.~G.}\ \bibnamefont
				{Makris}}, \bibinfo {author} {\bibfnamefont {M.}~\bibnamefont {Khajavikhan}},
			\bibinfo {author} {\bibfnamefont {Z.~H.}\ \bibnamefont {Musslimani}},
			\bibinfo {author} {\bibfnamefont {S.}~\bibnamefont {Rotter}},\ and\ \bibinfo
			{author} {\bibfnamefont {D.~N.}\ \bibnamefont {Christodoulides}},\ }\href
		{https://doi.org/10.1038/nphys4323} {\bibfield  {journal} {\bibinfo
				{journal} {Nature Phys}\ }\textbf {\bibinfo {volume} {14}},\ \bibinfo {pages}
			{11} (\bibinfo {year} {2018})}\BibitemShut {NoStop}%
		\bibitem [{\citenamefont {Wang}\ \emph {et~al.}(2021)\citenamefont {Wang},
			\citenamefont {Zhou}, \citenamefont {Zhang}, \citenamefont {Zhang},
			\citenamefont {Zhang}, \citenamefont {Xie}, \citenamefont {Wu}, \citenamefont
			{Chen}, \citenamefont {Ou}, \citenamefont {Wu}, \citenamefont {Jing},\ and\
			\citenamefont {Chen}}]{wang_observation_2021}%
		\BibitemOpen
		\bibfield  {author} {\bibinfo {author} {\bibfnamefont {W.-C.}\ \bibnamefont
				{Wang}}, \bibinfo {author} {\bibfnamefont {Y.-L.}\ \bibnamefont {Zhou}},
			\bibinfo {author} {\bibfnamefont {H.-L.}\ \bibnamefont {Zhang}}, \bibinfo
			{author} {\bibfnamefont {J.}~\bibnamefont {Zhang}}, \bibinfo {author}
			{\bibfnamefont {M.-C.}\ \bibnamefont {Zhang}}, \bibinfo {author}
			{\bibfnamefont {Y.}~\bibnamefont {Xie}}, \bibinfo {author} {\bibfnamefont
				{C.-W.}\ \bibnamefont {Wu}}, \bibinfo {author} {\bibfnamefont
				{T.}~\bibnamefont {Chen}}, \bibinfo {author} {\bibfnamefont {B.-Q.}\
				\bibnamefont {Ou}}, \bibinfo {author} {\bibfnamefont {W.}~\bibnamefont {Wu}},
			\bibinfo {author} {\bibfnamefont {H.}~\bibnamefont {Jing}},\ and\ \bibinfo
			{author} {\bibfnamefont {P.-X.}\ \bibnamefont {Chen}},\ }\href
		{https://doi.org/10.1103/PhysRevA.103.L020201} {\bibfield  {journal}
			{\bibinfo  {journal} {Phys. Rev. A}\ }\textbf {\bibinfo {volume} {103}},\
			\bibinfo {pages} {L020201} (\bibinfo {year} {2021})}\BibitemShut {NoStop}%
		\bibitem [{\citenamefont {Stehmann}\ \emph {et~al.}(2004)\citenamefont
			{Stehmann}, \citenamefont {Heiss},\ and\ \citenamefont
			{Scholtz}}]{T_Stehmann_2004}%
		\BibitemOpen
		\bibfield  {author} {\bibinfo {author} {\bibfnamefont {T.}~\bibnamefont
				{Stehmann}}, \bibinfo {author} {\bibfnamefont {W.~D.}\ \bibnamefont
				{Heiss}},\ and\ \bibinfo {author} {\bibfnamefont {F.~G.}\ \bibnamefont
				{Scholtz}},\ }\href {https://doi.org/10.1088/0305-4470/37/31/012} {\bibfield
			{journal} {\bibinfo  {journal} {Journal of Physics A: Mathematical and
					General}\ }\textbf {\bibinfo {volume} {37}},\ \bibinfo {pages} {7813}
			(\bibinfo {year} {2004})}\BibitemShut {NoStop}%
		\bibitem [{\citenamefont {Liertzer}\ \emph {et~al.}(2012)\citenamefont
			{Liertzer}, \citenamefont {Ge}, \citenamefont {Cerjan}, \citenamefont
			{Stone}, \citenamefont {Türeci},\ and\ \citenamefont
			{Rotter}}]{liertzer_pump-induced_2012}%
		\BibitemOpen
		\bibfield  {author} {\bibinfo {author} {\bibfnamefont {M.}~\bibnamefont
				{Liertzer}}, \bibinfo {author} {\bibfnamefont {L.}~\bibnamefont {Ge}},
			\bibinfo {author} {\bibfnamefont {A.}~\bibnamefont {Cerjan}}, \bibinfo
			{author} {\bibfnamefont {A.~D.}\ \bibnamefont {Stone}}, \bibinfo {author}
			{\bibfnamefont {H.~E.}\ \bibnamefont {Türeci}},\ and\ \bibinfo {author}
			{\bibfnamefont {S.}~\bibnamefont {Rotter}},\ }\href
		{https://doi.org/10.1103/PhysRevLett.108.173901} {\bibfield  {journal}
			{\bibinfo  {journal} {Phys. Rev. Lett.}\ }\textbf {\bibinfo {volume} {108}},\
			\bibinfo {pages} {173901} (\bibinfo {year} {2012})}\BibitemShut {NoStop}%
		\bibitem [{\citenamefont {Doppler}\ \emph {et~al.}(2016)\citenamefont
			{Doppler}, \citenamefont {Mailybaev}, \citenamefont {Böhm}, \citenamefont
			{Kuhl}, \citenamefont {Girschik}, \citenamefont {Libisch}, \citenamefont
			{Milburn}, \citenamefont {Rabl}, \citenamefont {Moiseyev},\ and\
			\citenamefont {Rotter}}]{doppler_dynamically_2016}%
		\BibitemOpen
		\bibfield  {author} {\bibinfo {author} {\bibfnamefont {J.}~\bibnamefont
				{Doppler}}, \bibinfo {author} {\bibfnamefont {A.~A.}\ \bibnamefont
				{Mailybaev}}, \bibinfo {author} {\bibfnamefont {J.}~\bibnamefont {Böhm}},
			\bibinfo {author} {\bibfnamefont {U.}~\bibnamefont {Kuhl}}, \bibinfo {author}
			{\bibfnamefont {A.}~\bibnamefont {Girschik}}, \bibinfo {author}
			{\bibfnamefont {F.}~\bibnamefont {Libisch}}, \bibinfo {author} {\bibfnamefont
				{T.~J.}\ \bibnamefont {Milburn}}, \bibinfo {author} {\bibfnamefont
				{P.}~\bibnamefont {Rabl}}, \bibinfo {author} {\bibfnamefont {N.}~\bibnamefont
				{Moiseyev}},\ and\ \bibinfo {author} {\bibfnamefont {S.}~\bibnamefont
				{Rotter}},\ }\href {https://doi.org/10.1038/nature18605} {\bibfield
			{journal} {\bibinfo  {journal} {Nature}\ }\textbf {\bibinfo {volume} {537}},\
			\bibinfo {pages} {76} (\bibinfo {year} {2016})}\BibitemShut {NoStop}%
		\bibitem [{\citenamefont {Li}\ \emph {et~al.}(2023)\citenamefont {Li},
			\citenamefont {Wei}, \citenamefont {Cotrufo}, \citenamefont {Chen},
			\citenamefont {Mann}, \citenamefont {Ni}, \citenamefont {Xu}, \citenamefont
			{Chen}, \citenamefont {Wang}, \citenamefont {Fan}, \citenamefont {Qiu},
			\citenamefont {Alù},\ and\ \citenamefont {Chen}}]{li_exceptional_2023}%
		\BibitemOpen
		\bibfield  {author} {\bibinfo {author} {\bibfnamefont {A.}~\bibnamefont
				{Li}}, \bibinfo {author} {\bibfnamefont {H.}~\bibnamefont {Wei}}, \bibinfo
			{author} {\bibfnamefont {M.}~\bibnamefont {Cotrufo}}, \bibinfo {author}
			{\bibfnamefont {W.}~\bibnamefont {Chen}}, \bibinfo {author} {\bibfnamefont
				{S.}~\bibnamefont {Mann}}, \bibinfo {author} {\bibfnamefont {X.}~\bibnamefont
				{Ni}}, \bibinfo {author} {\bibfnamefont {B.}~\bibnamefont {Xu}}, \bibinfo
			{author} {\bibfnamefont {J.}~\bibnamefont {Chen}}, \bibinfo {author}
			{\bibfnamefont {J.}~\bibnamefont {Wang}}, \bibinfo {author} {\bibfnamefont
				{S.}~\bibnamefont {Fan}}, \bibinfo {author} {\bibfnamefont {C.-W.}\
				\bibnamefont {Qiu}}, \bibinfo {author} {\bibfnamefont {A.}~\bibnamefont
				{Alù}},\ and\ \bibinfo {author} {\bibfnamefont {L.}~\bibnamefont {Chen}},\
		}\href {https://doi.org/10.1038/s41565-023-01408-0} {\bibfield  {journal}
			{\bibinfo  {journal} {Nat. Nanotechnol.}\ }\textbf {\bibinfo {volume} {18}},\
			\bibinfo {pages} {706720} (\bibinfo {year} {2023})}\BibitemShut {NoStop}%
		\bibitem [{\citenamefont {Rabi}(1936)}]{rabi_process_1936}%
		\BibitemOpen
		\bibfield  {author} {\bibinfo {author} {\bibfnamefont {I.~I.}\ \bibnamefont
				{Rabi}},\ }\href {https://doi.org/10.1103/PhysRev.49.324} {\bibfield
			{journal} {\bibinfo  {journal} {Phys. Rev.}\ }\textbf {\bibinfo {volume}
				{49}},\ \bibinfo {pages} {324} (\bibinfo {year} {1936})}\BibitemShut
		{NoStop}%
		\bibitem [{\citenamefont {Braak}(2011)}]{braak_integrability_2011}%
		\BibitemOpen
		\bibfield  {author} {\bibinfo {author} {\bibfnamefont {D.}~\bibnamefont
				{Braak}},\ }\href {https://doi.org/10.1103/PhysRevLett.107.100401} {\bibfield
			{journal} {\bibinfo  {journal} {Phys. Rev. Lett.}\ }\textbf {\bibinfo
				{volume} {107}},\ \bibinfo {pages} {100401} (\bibinfo {year}
			{2011})}\BibitemShut {NoStop}%
		\bibitem [{\citenamefont {Scully}\ and\ \citenamefont
			{Zubairy}(1997)}]{scully_zubairy_1997}%
		\BibitemOpen
		\bibfield  {author} {\bibinfo {author} {\bibfnamefont {M.~O.}\ \bibnamefont
				{Scully}}\ and\ \bibinfo {author} {\bibfnamefont {M.~S.}\ \bibnamefont
				{Zubairy}},\ }\href {https://doi.org/10.1017/CBO9780511813993} {\emph
			{\bibinfo {title} {Quantum Optics}}}\ (\bibinfo  {publisher} {Cambridge
			University Press},\ \bibinfo {year} {1997})\BibitemShut {NoStop}%
		\bibitem [{\citenamefont {Braak}\ \emph {et~al.}(2016)\citenamefont {Braak},
			\citenamefont {Chen}, \citenamefont {Batchelor},\ and\ \citenamefont
			{Solano}}]{braak_semi-classical_2016}%
		\BibitemOpen
		\bibfield  {author} {\bibinfo {author} {\bibfnamefont {D.}~\bibnamefont
				{Braak}}, \bibinfo {author} {\bibfnamefont {Q.-H.}\ \bibnamefont {Chen}},
			\bibinfo {author} {\bibfnamefont {M.~T.}\ \bibnamefont {Batchelor}},\ and\
			\bibinfo {author} {\bibfnamefont {E.}~\bibnamefont {Solano}},\ }\href
		{https://doi.org/10.1088/1751-8113/49/30/300301} {\bibfield  {journal}
			{\bibinfo  {journal} {J. Phys. A: Math. Theor.}\ }\textbf {\bibinfo {volume}
				{49}},\ \bibinfo {pages} {300301} (\bibinfo {year} {2016})}\BibitemShut
		{NoStop}%
		\bibitem [{\citenamefont {Meystre}(2021)}]{meystre_quantum_2021}%
		\BibitemOpen
		\bibfield  {author} {\bibinfo {author} {\bibfnamefont {P.}~\bibnamefont
				{Meystre}},\ }\href {https://doi.org/10.1007/978-3-030-76183-7} {\emph
			{\bibinfo {title} {Quantum {Optics}: {Taming} the {Quantum}}}},\ Graduate
		{Texts} in {Physics}\ (\bibinfo  {publisher} {Springer},\ \bibinfo {address}
		{Berlin},\ \bibinfo {year} {2021})\BibitemShut {NoStop}%
		\bibitem [{\citenamefont {Niemczyk}\ \emph {et~al.}(2010)\citenamefont
			{Niemczyk}, \citenamefont {Deppe}, \citenamefont {Huebl}, \citenamefont
			{Menzel}, \citenamefont {Hocke}, \citenamefont {Schwarz}, \citenamefont
			{Garcia-Ripoll}, \citenamefont {Zueco}, \citenamefont {Hümmer},
			\citenamefont {Solano}, \citenamefont {Marx},\ and\ \citenamefont
			{Gross}}]{niemczyk_circuit_2010}%
		\BibitemOpen
		\bibfield  {author} {\bibinfo {author} {\bibfnamefont {T.}~\bibnamefont
				{Niemczyk}}, \bibinfo {author} {\bibfnamefont {F.}~\bibnamefont {Deppe}},
			\bibinfo {author} {\bibfnamefont {H.}~\bibnamefont {Huebl}}, \bibinfo
			{author} {\bibfnamefont {E.~P.}\ \bibnamefont {Menzel}}, \bibinfo {author}
			{\bibfnamefont {F.}~\bibnamefont {Hocke}}, \bibinfo {author} {\bibfnamefont
				{M.~J.}\ \bibnamefont {Schwarz}}, \bibinfo {author} {\bibfnamefont {J.~J.}\
				\bibnamefont {Garcia-Ripoll}}, \bibinfo {author} {\bibfnamefont
				{D.}~\bibnamefont {Zueco}}, \bibinfo {author} {\bibfnamefont
				{T.}~\bibnamefont {Hümmer}}, \bibinfo {author} {\bibfnamefont
				{E.}~\bibnamefont {Solano}}, \bibinfo {author} {\bibfnamefont
				{A.}~\bibnamefont {Marx}},\ and\ \bibinfo {author} {\bibfnamefont
				{R.}~\bibnamefont {Gross}},\ }\href {https://doi.org/10.1038/nphys1730}
		{\bibfield  {journal} {\bibinfo  {journal} {Nature Phys}\ }\textbf {\bibinfo
				{volume} {6}},\ \bibinfo {pages} {772} (\bibinfo {year} {2010})}\BibitemShut
		{NoStop}%
		\bibitem [{\citenamefont {Forn-Díaz}\ \emph {et~al.}(2017)\citenamefont
			{Forn-Díaz}, \citenamefont {García-Ripoll}, \citenamefont {Peropadre},
			\citenamefont {Orgiazzi}, \citenamefont {Yurtalan}, \citenamefont
			{Belyansky}, \citenamefont {Wilson},\ and\ \citenamefont
			{Lupascu}}]{forn-diaz_ultrastrong_2017}%
		\BibitemOpen
		\bibfield  {author} {\bibinfo {author} {\bibfnamefont {P.}~\bibnamefont
				{Forn-Díaz}}, \bibinfo {author} {\bibfnamefont {J.~J.}\ \bibnamefont
				{García-Ripoll}}, \bibinfo {author} {\bibfnamefont {B.}~\bibnamefont
				{Peropadre}}, \bibinfo {author} {\bibfnamefont {J.-L.}\ \bibnamefont
				{Orgiazzi}}, \bibinfo {author} {\bibfnamefont {M.~A.}\ \bibnamefont
				{Yurtalan}}, \bibinfo {author} {\bibfnamefont {R.}~\bibnamefont {Belyansky}},
			\bibinfo {author} {\bibfnamefont {C.~M.}\ \bibnamefont {Wilson}},\ and\
			\bibinfo {author} {\bibfnamefont {A.}~\bibnamefont {Lupascu}},\ }\href
		{https://doi.org/10.1038/nphys3905} {\bibfield  {journal} {\bibinfo
				{journal} {Nature Phys}\ }\textbf {\bibinfo {volume} {13}},\ \bibinfo {pages}
			{39} (\bibinfo {year} {2017})}\BibitemShut {NoStop}%
		\bibitem [{\citenamefont {Forn-Díaz}\ \emph {et~al.}(2019)\citenamefont
			{Forn-Díaz}, \citenamefont {Lamata}, \citenamefont {Rico}, \citenamefont
			{Kono},\ and\ \citenamefont {Solano}}]{forn-diaz_ultrastrong_2019}%
		\BibitemOpen
		\bibfield  {author} {\bibinfo {author} {\bibfnamefont {P.}~\bibnamefont
				{Forn-Díaz}}, \bibinfo {author} {\bibfnamefont {L.}~\bibnamefont {Lamata}},
			\bibinfo {author} {\bibfnamefont {E.}~\bibnamefont {Rico}}, \bibinfo {author}
			{\bibfnamefont {J.}~\bibnamefont {Kono}},\ and\ \bibinfo {author}
			{\bibfnamefont {E.}~\bibnamefont {Solano}},\ }\href
		{https://doi.org/10.1103/RevModPhys.91.025005} {\bibfield  {journal}
			{\bibinfo  {journal} {Rev. Mod. Phys.}\ }\textbf {\bibinfo {volume} {91}},\
			\bibinfo {pages} {025005} (\bibinfo {year} {2019})}\BibitemShut {NoStop}%
		\bibitem [{\citenamefont {Leibfried}\ \emph {et~al.}(2003)\citenamefont
			{Leibfried}, \citenamefont {Blatt}, \citenamefont {Monroe},\ and\
			\citenamefont {Wineland}}]{leibfried_quantum_2003}%
		\BibitemOpen
		\bibfield  {author} {\bibinfo {author} {\bibfnamefont {D.}~\bibnamefont
				{Leibfried}}, \bibinfo {author} {\bibfnamefont {R.}~\bibnamefont {Blatt}},
			\bibinfo {author} {\bibfnamefont {C.}~\bibnamefont {Monroe}},\ and\ \bibinfo
			{author} {\bibfnamefont {D.}~\bibnamefont {Wineland}},\ }\href
		{https://doi.org/10.1103/RevModPhys.75.281} {\bibfield  {journal} {\bibinfo
				{journal} {Rev. Mod. Phys.}\ }\textbf {\bibinfo {volume} {75}},\ \bibinfo
			{pages} {281} (\bibinfo {year} {2003})}\BibitemShut {NoStop}%
		\bibitem [{\citenamefont {Pedernales}\ \emph {et~al.}(2015)\citenamefont
			{Pedernales}, \citenamefont {Lizuain}, \citenamefont {Felicetti},
			\citenamefont {Romero}, \citenamefont {Lamata},\ and\ \citenamefont
			{Solano}}]{pedernales_quantum_2015}%
		\BibitemOpen
		\bibfield  {author} {\bibinfo {author} {\bibfnamefont {J.~S.}\ \bibnamefont
				{Pedernales}}, \bibinfo {author} {\bibfnamefont {I.}~\bibnamefont {Lizuain}},
			\bibinfo {author} {\bibfnamefont {S.}~\bibnamefont {Felicetti}}, \bibinfo
			{author} {\bibfnamefont {G.}~\bibnamefont {Romero}}, \bibinfo {author}
			{\bibfnamefont {L.}~\bibnamefont {Lamata}},\ and\ \bibinfo {author}
			{\bibfnamefont {E.}~\bibnamefont {Solano}},\ }\href
		{https://doi.org/10.1038/srep15472} {\bibfield  {journal} {\bibinfo
				{journal} {Sci Rep}\ }\textbf {\bibinfo {volume} {5}},\ \bibinfo {pages}
			{15472} (\bibinfo {year} {2015})}\BibitemShut {NoStop}%
		\bibitem [{\citenamefont {Cai}\ \emph {et~al.}(2021)\citenamefont {Cai},
			\citenamefont {Liu}, \citenamefont {Zhao}, \citenamefont {Wu}, \citenamefont
			{Mei}, \citenamefont {Jiang}, \citenamefont {He}, \citenamefont {Zhang},
			\citenamefont {Zhou},\ and\ \citenamefont {Duan}}]{cai_observation_2021}%
		\BibitemOpen
		\bibfield  {author} {\bibinfo {author} {\bibfnamefont {M.-L.}\ \bibnamefont
				{Cai}}, \bibinfo {author} {\bibfnamefont {Z.-D.}\ \bibnamefont {Liu}},
			\bibinfo {author} {\bibfnamefont {W.-D.}\ \bibnamefont {Zhao}}, \bibinfo
			{author} {\bibfnamefont {Y.-K.}\ \bibnamefont {Wu}}, \bibinfo {author}
			{\bibfnamefont {Q.-X.}\ \bibnamefont {Mei}}, \bibinfo {author} {\bibfnamefont
				{Y.}~\bibnamefont {Jiang}}, \bibinfo {author} {\bibfnamefont
				{L.}~\bibnamefont {He}}, \bibinfo {author} {\bibfnamefont {X.}~\bibnamefont
				{Zhang}}, \bibinfo {author} {\bibfnamefont {Z.-C.}\ \bibnamefont {Zhou}},\
			and\ \bibinfo {author} {\bibfnamefont {L.-M.}\ \bibnamefont {Duan}},\ }\href
		{https://doi.org/10.1038/s41467-021-21425-8} {\bibfield  {journal} {\bibinfo
				{journal} {Nat Commun}\ }\textbf {\bibinfo {volume} {12}},\ \bibinfo {pages}
			{1126} (\bibinfo {year} {2021})}\BibitemShut {NoStop}%
		\bibitem [{\citenamefont {Bertet}\ \emph {et~al.}(2002)\citenamefont {Bertet},
			\citenamefont {Osnaghi}, \citenamefont {Milman}, \citenamefont {Auffeves},
			\citenamefont {Maioli}, \citenamefont {Brune}, \citenamefont {Raimond},\ and\
			\citenamefont {Haroche}}]{bertet_generating_2002}%
		\BibitemOpen
		\bibfield  {author} {\bibinfo {author} {\bibfnamefont {P.}~\bibnamefont
				{Bertet}}, \bibinfo {author} {\bibfnamefont {S.}~\bibnamefont {Osnaghi}},
			\bibinfo {author} {\bibfnamefont {P.}~\bibnamefont {Milman}}, \bibinfo
			{author} {\bibfnamefont {A.}~\bibnamefont {Auffeves}}, \bibinfo {author}
			{\bibfnamefont {P.}~\bibnamefont {Maioli}}, \bibinfo {author} {\bibfnamefont
				{M.}~\bibnamefont {Brune}}, \bibinfo {author} {\bibfnamefont {J.~M.}\
				\bibnamefont {Raimond}},\ and\ \bibinfo {author} {\bibfnamefont
				{S.}~\bibnamefont {Haroche}},\ }\href
		{https://doi.org/10.1103/PhysRevLett.88.143601} {\bibfield  {journal}
			{\bibinfo  {journal} {Phys. Rev. Lett.}\ }\textbf {\bibinfo {volume} {88}},\
			\bibinfo {pages} {143601} (\bibinfo {year} {2002})}\BibitemShut {NoStop}%
		\bibitem [{\citenamefont {Stufler}\ \emph {et~al.}(2006)\citenamefont
			{Stufler}, \citenamefont {Machnikowski}, \citenamefont {Ester}, \citenamefont
			{Bichler}, \citenamefont {Axt}, \citenamefont {Kuhn},\ and\ \citenamefont
			{Zrenner}}]{stufler_two-photon_2006}%
		\BibitemOpen
		\bibfield  {author} {\bibinfo {author} {\bibfnamefont {S.}~\bibnamefont
				{Stufler}}, \bibinfo {author} {\bibfnamefont {P.}~\bibnamefont
				{Machnikowski}}, \bibinfo {author} {\bibfnamefont {P.}~\bibnamefont {Ester}},
			\bibinfo {author} {\bibfnamefont {M.}~\bibnamefont {Bichler}}, \bibinfo
			{author} {\bibfnamefont {V.~M.}\ \bibnamefont {Axt}}, \bibinfo {author}
			{\bibfnamefont {T.}~\bibnamefont {Kuhn}},\ and\ \bibinfo {author}
			{\bibfnamefont {A.}~\bibnamefont {Zrenner}},\ }\href
		{https://doi.org/10.1103/PhysRevB.73.125304} {\bibfield  {journal} {\bibinfo
				{journal} {Phys. Rev. B}\ }\textbf {\bibinfo {volume} {73}},\ \bibinfo
			{pages} {125304} (\bibinfo {year} {2006})}\BibitemShut {NoStop}%
		\bibitem [{\citenamefont {Del~Valle}\ \emph {et~al.}(2010)\citenamefont
			{Del~Valle}, \citenamefont {Zippilli}, \citenamefont {Laussy}, \citenamefont
			{Gonzalez-Tudela}, \citenamefont {Morigi},\ and\ \citenamefont
			{Tejedor}}]{del_valle_two-photon_2010}%
		\BibitemOpen
		\bibfield  {author} {\bibinfo {author} {\bibfnamefont {E.}~\bibnamefont
				{Del~Valle}}, \bibinfo {author} {\bibfnamefont {S.}~\bibnamefont {Zippilli}},
			\bibinfo {author} {\bibfnamefont {F.~P.}\ \bibnamefont {Laussy}}, \bibinfo
			{author} {\bibfnamefont {A.}~\bibnamefont {Gonzalez-Tudela}}, \bibinfo
			{author} {\bibfnamefont {G.}~\bibnamefont {Morigi}},\ and\ \bibinfo {author}
			{\bibfnamefont {C.}~\bibnamefont {Tejedor}},\ }\href
		{https://doi.org/10.1103/PhysRevB.81.035302} {\bibfield  {journal} {\bibinfo
				{journal} {Phys. Rev. B}\ }\textbf {\bibinfo {volume} {81}},\ \bibinfo
			{pages} {035302} (\bibinfo {year} {2010})}\BibitemShut {NoStop}%
		\bibitem [{\citenamefont {Cong}\ \emph {et~al.}(2019)\citenamefont {Cong},
			\citenamefont {Sun}, \citenamefont {Liu}, \citenamefont {Ying},\ and\
			\citenamefont {Luo}}]{cong_polaron_2019}%
		\BibitemOpen
		\bibfield  {author} {\bibinfo {author} {\bibfnamefont {L.}~\bibnamefont
				{Cong}}, \bibinfo {author} {\bibfnamefont {X.-M.}\ \bibnamefont {Sun}},
			\bibinfo {author} {\bibfnamefont {M.}~\bibnamefont {Liu}}, \bibinfo {author}
			{\bibfnamefont {Z.-J.}\ \bibnamefont {Ying}},\ and\ \bibinfo {author}
			{\bibfnamefont {H.-G.}\ \bibnamefont {Luo}},\ }\href
		{https://doi.org/10.1103/PhysRevA.99.013815} {\bibfield  {journal} {\bibinfo
				{journal} {Phys. Rev. A}\ }\textbf {\bibinfo {volume} {99}},\ \bibinfo
			{pages} {013815} (\bibinfo {year} {2019})}\BibitemShut {NoStop}%
		\bibitem [{\citenamefont {Felicetti}\ \emph {et~al.}(2015)\citenamefont
			{Felicetti}, \citenamefont {Pedernales}, \citenamefont {Egusquiza},
			\citenamefont {Romero}, \citenamefont {Lamata}, \citenamefont {Braak},\ and\
			\citenamefont {Solano}}]{felicetti_spectral_2015}%
		\BibitemOpen
		\bibfield  {author} {\bibinfo {author} {\bibfnamefont {S.}~\bibnamefont
				{Felicetti}}, \bibinfo {author} {\bibfnamefont {J.~S.}\ \bibnamefont
				{Pedernales}}, \bibinfo {author} {\bibfnamefont {I.~L.}\ \bibnamefont
				{Egusquiza}}, \bibinfo {author} {\bibfnamefont {G.}~\bibnamefont {Romero}},
			\bibinfo {author} {\bibfnamefont {L.}~\bibnamefont {Lamata}}, \bibinfo
			{author} {\bibfnamefont {D.}~\bibnamefont {Braak}},\ and\ \bibinfo {author}
			{\bibfnamefont {E.}~\bibnamefont {Solano}},\ }\href
		{https://doi.org/10.1103/PhysRevA.92.033817} {\bibfield  {journal} {\bibinfo
				{journal} {Phys. Rev. A}\ }\textbf {\bibinfo {volume} {92}},\ \bibinfo
			{pages} {033817} (\bibinfo {year} {2015})}\BibitemShut {NoStop}%
		\bibitem [{\citenamefont {Duan}\ \emph {et~al.}(2016)\citenamefont {Duan},
			\citenamefont {Xie}, \citenamefont {Braak},\ and\ \citenamefont
			{Chen}}]{duan_two-photon_2016}%
		\BibitemOpen
		\bibfield  {author} {\bibinfo {author} {\bibfnamefont {L.}~\bibnamefont
				{Duan}}, \bibinfo {author} {\bibfnamefont {Y.-F.}\ \bibnamefont {Xie}},
			\bibinfo {author} {\bibfnamefont {D.}~\bibnamefont {Braak}},\ and\ \bibinfo
			{author} {\bibfnamefont {Q.-H.}\ \bibnamefont {Chen}},\ }\href
		{https://doi.org/10.1088/1751-8113/49/46/464002} {\bibfield  {journal}
			{\bibinfo  {journal} {J. Phys. A: Math. Theor.}\ }\textbf {\bibinfo {volume}
				{49}},\ \bibinfo {pages} {464002} (\bibinfo {year} {2016})}\BibitemShut
		{NoStop}%
		\bibitem [{\citenamefont {Ying}(2021)}]{ying_symmetry-breaking_2021}%
		\BibitemOpen
		\bibfield  {author} {\bibinfo {author} {\bibfnamefont {Z.-J.}\ \bibnamefont
				{Ying}},\ }\href {https://doi.org/10.1103/PhysRevA.103.063701} {\bibfield
			{journal} {\bibinfo  {journal} {Phys. Rev. A}\ }\textbf {\bibinfo {volume}
				{103}},\ \bibinfo {pages} {063701} (\bibinfo {year} {2021})}\BibitemShut
		{NoStop}%
		\bibitem [{\citenamefont {Ben-Aryeh}\ \emph {et~al.}(2004)\citenamefont
			{Ben-Aryeh}, \citenamefont {Mann},\ and\ \citenamefont
			{Yaakov}}]{Ben-Aryeh_2004}%
		\BibitemOpen
		\bibfield  {author} {\bibinfo {author} {\bibfnamefont {Y.}~\bibnamefont
				{Ben-Aryeh}}, \bibinfo {author} {\bibfnamefont {A.}~\bibnamefont {Mann}},\
			and\ \bibinfo {author} {\bibfnamefont {I.}~\bibnamefont {Yaakov}},\ }\href
		{https://doi.org/10.1088/0305-4470/37/50/008} {\bibfield  {journal} {\bibinfo
				{journal} {Journal of Physics A: Mathematical and General}\ }\textbf
			{\bibinfo {volume} {37}},\ \bibinfo {pages} {12059} (\bibinfo {year}
			{2004})}\BibitemShut {NoStop}%
		\bibitem [{\citenamefont {Joglekar}\ \emph {et~al.}(2014)\citenamefont
			{Joglekar}, \citenamefont {Marathe}, \citenamefont {Durganandini},\ and\
			\citenamefont {Pathak}}]{joglekar_pt_2014}%
		\BibitemOpen
		\bibfield  {author} {\bibinfo {author} {\bibfnamefont {Y.~N.}\ \bibnamefont
				{Joglekar}}, \bibinfo {author} {\bibfnamefont {R.}~\bibnamefont {Marathe}},
			\bibinfo {author} {\bibfnamefont {P.}~\bibnamefont {Durganandini}},\ and\
			\bibinfo {author} {\bibfnamefont {R.~K.}\ \bibnamefont {Pathak}},\ }\href
		{https://doi.org/10.1103/PhysRevA.90.040101} {\bibfield  {journal} {\bibinfo
				{journal} {Phys. Rev. A}\ }\textbf {\bibinfo {volume} {90}},\ \bibinfo
			{pages} {040101} (\bibinfo {year} {2014})}\BibitemShut {NoStop}%
		\bibitem [{\citenamefont {Lee}\ and\ \citenamefont
			{Joglekar}(2015)}]{lee_pt_2015}%
		\BibitemOpen
		\bibfield  {author} {\bibinfo {author} {\bibfnamefont {T.~E.}\ \bibnamefont
				{Lee}}\ and\ \bibinfo {author} {\bibfnamefont {Y.~N.}\ \bibnamefont
				{Joglekar}},\ }\href {https://doi.org/10.1103/PhysRevA.92.042103} {\bibfield
			{journal} {\bibinfo  {journal} {Phys. Rev. A}\ }\textbf {\bibinfo {volume}
				{92}},\ \bibinfo {pages} {042103} (\bibinfo {year} {2015})}\BibitemShut
		{NoStop}%
		\bibitem [{\citenamefont {Xie}\ \emph {et~al.}(2018)\citenamefont {Xie},
			\citenamefont {Rong},\ and\ \citenamefont {Liu}}]{xie_exceptional_2018}%
		\BibitemOpen
		\bibfield  {author} {\bibinfo {author} {\bibfnamefont {Q.}~\bibnamefont
				{Xie}}, \bibinfo {author} {\bibfnamefont {S.}~\bibnamefont {Rong}},\ and\
			\bibinfo {author} {\bibfnamefont {X.}~\bibnamefont {Liu}},\ }\href
		{https://doi.org/10.1103/PhysRevA.98.052122} {\bibfield  {journal} {\bibinfo
				{journal} {Phys. Rev. A}\ }\textbf {\bibinfo {volume} {98}},\ \bibinfo
			{pages} {052122} (\bibinfo {year} {2018})}\BibitemShut {NoStop}%
		\bibitem [{\citenamefont {Lu}\ \emph {et~al.}(2023)\citenamefont {Lu},
			\citenamefont {Li}, \citenamefont {Shi}, \citenamefont {Fan}, \citenamefont
			{Mangazeev}, \citenamefont {Li},\ and\ \citenamefont
			{Batchelor}}]{lu_pt_2023}%
		\BibitemOpen
		\bibfield  {author} {\bibinfo {author} {\bibfnamefont {X.}~\bibnamefont
				{Lu}}, \bibinfo {author} {\bibfnamefont {H.}~\bibnamefont {Li}}, \bibinfo
			{author} {\bibfnamefont {J.-K.}\ \bibnamefont {Shi}}, \bibinfo {author}
			{\bibfnamefont {L.-B.}\ \bibnamefont {Fan}}, \bibinfo {author} {\bibfnamefont
				{V.}~\bibnamefont {Mangazeev}}, \bibinfo {author} {\bibfnamefont {Z.-M.}\
				\bibnamefont {Li}},\ and\ \bibinfo {author} {\bibfnamefont {M.~T.}\
				\bibnamefont {Batchelor}},\ }\href
		{https://doi.org/10.1103/PhysRevA.108.053712} {\bibfield  {journal} {\bibinfo
				{journal} {Phys. Rev. A}\ }\textbf {\bibinfo {volume} {108}},\ \bibinfo
			{pages} {053712} (\bibinfo {year} {2023})}\BibitemShut {NoStop}%
		\bibitem [{\citenamefont {Luo}\ \emph {et~al.}(2024)\citenamefont {Luo},
			\citenamefont {Liu},\ and\ \citenamefont {Liang}}]{luo_quantum_2024}%
		\BibitemOpen
		\bibfield  {author} {\bibinfo {author} {\bibfnamefont {Y.}~\bibnamefont
				{Luo}}, \bibinfo {author} {\bibfnamefont {N.}~\bibnamefont {Liu}},\ and\
			\bibinfo {author} {\bibfnamefont {J.-Q.}\ \bibnamefont {Liang}},\ }\href
		{https://doi.org/10.1103/PhysRevA.110.063320} {\bibfield  {journal} {\bibinfo
				{journal} {Phys. Rev. A}\ }\textbf {\bibinfo {volume} {110}},\ \bibinfo
			{pages} {063320} (\bibinfo {year} {2024})}\BibitemShut {NoStop}%
		\bibitem [{\citenamefont {Li}\ \emph {et~al.}(2025)\citenamefont {Li},
			\citenamefont {Wang}, \citenamefont {Duan},\ and\ \citenamefont
			{Chen}}]{li_ptmathcal_2025}%
		\BibitemOpen
		\bibfield  {author} {\bibinfo {author} {\bibfnamefont {J.}~\bibnamefont
				{Li}}, \bibinfo {author} {\bibfnamefont {Y.}~\bibnamefont {Wang}}, \bibinfo
			{author} {\bibfnamefont {L.}~\bibnamefont {Duan}},\ and\ \bibinfo {author}
			{\bibfnamefont {Q.}~\bibnamefont {Chen}},\ }\href
		{https://doi.org/10.1002/qute.202400609} {\bibfield  {journal} {\bibinfo
				{journal} {Adv Quantum Tech}\ ,\ \bibinfo {pages} {2400609}} (\bibinfo {year}
			{2025})}\BibitemShut {NoStop}%
		\bibitem [{\citenamefont {Chen}\ \emph {et~al.}(2012)\citenamefont {Chen},
			\citenamefont {Wang}, \citenamefont {He}, \citenamefont {Liu},\ and\
			\citenamefont {Wang}}]{chen_exact_2012}%
		\BibitemOpen
		\bibfield  {author} {\bibinfo {author} {\bibfnamefont {Q.-H.}\ \bibnamefont
				{Chen}}, \bibinfo {author} {\bibfnamefont {C.}~\bibnamefont {Wang}}, \bibinfo
			{author} {\bibfnamefont {S.}~\bibnamefont {He}}, \bibinfo {author}
			{\bibfnamefont {T.}~\bibnamefont {Liu}},\ and\ \bibinfo {author}
			{\bibfnamefont {K.-L.}\ \bibnamefont {Wang}},\ }\href
		{https://doi.org/10.1103/PhysRevA.86.023822} {\bibfield  {journal} {\bibinfo
				{journal} {Phys. Rev. A}\ }\textbf {\bibinfo {volume} {86}},\ \bibinfo
			{pages} {023822} (\bibinfo {year} {2012})}\BibitemShut {NoStop}%
		\bibitem [{\citenamefont {Xie}\ \emph {et~al.}(2017)\citenamefont {Xie},
			\citenamefont {Zhong}, \citenamefont {Batchelor},\ and\ \citenamefont
			{Lee}}]{xie_quantum_2017}%
		\BibitemOpen
		\bibfield  {author} {\bibinfo {author} {\bibfnamefont {Q.}~\bibnamefont
				{Xie}}, \bibinfo {author} {\bibfnamefont {H.}~\bibnamefont {Zhong}}, \bibinfo
			{author} {\bibfnamefont {M.~T.}\ \bibnamefont {Batchelor}},\ and\ \bibinfo
			{author} {\bibfnamefont {C.}~\bibnamefont {Lee}},\ }\href
		{https://doi.org/10.1088/1751-8121/aa5a65} {\bibfield  {journal} {\bibinfo
				{journal} {J. Phys. A: Math. Theor.}\ }\textbf {\bibinfo {volume} {50}},\
			\bibinfo {pages} {113001} (\bibinfo {year} {2017})}\BibitemShut {NoStop}%
		\bibitem [{\citenamefont {Li}\ and\ \citenamefont
			{Chen}(2020)}]{li_two-photon_2020}%
		\BibitemOpen
		\bibfield  {author} {\bibinfo {author} {\bibfnamefont {J.}~\bibnamefont
				{Li}}\ and\ \bibinfo {author} {\bibfnamefont {Q.-H.}\ \bibnamefont {Chen}},\
		}\href {https://doi.org/10.1088/1751-8121/ab8ef1} {\bibfield  {journal}
			{\bibinfo  {journal} {J. Phys. A: Math. Theor.}\ }\textbf {\bibinfo {volume}
				{53}},\ \bibinfo {pages} {315301} (\bibinfo {year} {2020})}\BibitemShut
		{NoStop}%
		\bibitem [{\citenamefont {Xie}\ and\ \citenamefont
			{Chen}(2021)}]{xie_double_2021}%
		\BibitemOpen
		\bibfield  {author} {\bibinfo {author} {\bibfnamefont {Y.-F.}\ \bibnamefont
				{Xie}}\ and\ \bibinfo {author} {\bibfnamefont {Q.-H.}\ \bibnamefont {Chen}},\
		}\href {https://doi.org/10.1103/PhysRevResearch.3.033057} {\bibfield
			{journal} {\bibinfo  {journal} {Phys. Rev. Research}\ }\textbf {\bibinfo
				{volume} {3}},\ \bibinfo {pages} {033057} (\bibinfo {year}
			{2021})}\BibitemShut {NoStop}%
		\bibitem [{\citenamefont {Duan}(2022)}]{duan_unified_2022}%
		\BibitemOpen
		\bibfield  {author} {\bibinfo {author} {\bibfnamefont {L.}~\bibnamefont
				{Duan}},\ }\href {https://doi.org/10.1088/1367-2630/ac8a68} {\bibfield
			{journal} {\bibinfo  {journal} {New J. Phys.}\ }\textbf {\bibinfo {volume}
				{24}},\ \bibinfo {pages} {083045} (\bibinfo {year} {2022})}\BibitemShut
		{NoStop}%
		\bibitem [{\citenamefont {Braak}(2023)}]{braak_spectral_2023}%
		\BibitemOpen
		\bibfield  {author} {\bibinfo {author} {\bibfnamefont {D.}~\bibnamefont
				{Braak}},\ }\href {https://doi.org/10.1002/andp.202200519} {\bibfield
			{journal} {\bibinfo  {journal} {Annalen der Physik}\ }\textbf {\bibinfo
				{volume} {535}},\ \bibinfo {pages} {2200519} (\bibinfo {year}
			{2023})}\BibitemShut {NoStop}%
		\bibitem [{\citenamefont {Felicetti}\ \emph {et~al.}(2018)\citenamefont
			{Felicetti}, \citenamefont {Rossatto}, \citenamefont {Rico}, \citenamefont
			{Solano},\ and\ \citenamefont {Forn-Díaz}}]{felicetti_two-photon_2018}%
		\BibitemOpen
		\bibfield  {author} {\bibinfo {author} {\bibfnamefont {S.}~\bibnamefont
				{Felicetti}}, \bibinfo {author} {\bibfnamefont {D.~Z.}\ \bibnamefont
				{Rossatto}}, \bibinfo {author} {\bibfnamefont {E.}~\bibnamefont {Rico}},
			\bibinfo {author} {\bibfnamefont {E.}~\bibnamefont {Solano}},\ and\ \bibinfo
			{author} {\bibfnamefont {P.}~\bibnamefont {Forn-Díaz}},\ }\href
		{https://doi.org/10.1103/PhysRevA.97.013851} {\bibfield  {journal} {\bibinfo
				{journal} {Phys. Rev. A}\ }\textbf {\bibinfo {volume} {97}},\ \bibinfo
			{pages} {013851} (\bibinfo {year} {2018})}\BibitemShut {NoStop}%
		\bibitem [{\citenamefont {Naghiloo}\ \emph {et~al.}(2019)\citenamefont
			{Naghiloo}, \citenamefont {Abbasi}, \citenamefont {Joglekar},\ and\
			\citenamefont {Murch}}]{naghiloo_quantum_2019}%
		\BibitemOpen
		\bibfield  {author} {\bibinfo {author} {\bibfnamefont {M.}~\bibnamefont
				{Naghiloo}}, \bibinfo {author} {\bibfnamefont {M.}~\bibnamefont {Abbasi}},
			\bibinfo {author} {\bibfnamefont {Y.~N.}\ \bibnamefont {Joglekar}},\ and\
			\bibinfo {author} {\bibfnamefont {K.~W.}\ \bibnamefont {Murch}},\ }\href
		{https://doi.org/10.1038/s41567-019-0652-z} {\bibfield  {journal} {\bibinfo
				{journal} {Nat. Phys.}\ }\textbf {\bibinfo {volume} {15}},\ \bibinfo {pages}
			{1232} (\bibinfo {year} {2019})}\BibitemShut {NoStop}%
		\bibitem [{\citenamefont {Peng}\ \emph {et~al.}(2016)\citenamefont {Peng},
			\citenamefont {Cao}, \citenamefont {Shen}, \citenamefont {Qu}, \citenamefont
			{Wen}, \citenamefont {Jiang},\ and\ \citenamefont
			{Xiao}}]{peng_anti-paritytime_2016}%
		\BibitemOpen
		\bibfield  {author} {\bibinfo {author} {\bibfnamefont {P.}~\bibnamefont
				{Peng}}, \bibinfo {author} {\bibfnamefont {W.}~\bibnamefont {Cao}}, \bibinfo
			{author} {\bibfnamefont {C.}~\bibnamefont {Shen}}, \bibinfo {author}
			{\bibfnamefont {W.}~\bibnamefont {Qu}}, \bibinfo {author} {\bibfnamefont
				{J.}~\bibnamefont {Wen}}, \bibinfo {author} {\bibfnamefont {L.}~\bibnamefont
				{Jiang}},\ and\ \bibinfo {author} {\bibfnamefont {Y.}~\bibnamefont {Xiao}},\
		}\href {https://doi.org/10.1038/nphys3842} {\bibfield  {journal} {\bibinfo
				{journal} {Nature Phys}\ }\textbf {\bibinfo {volume} {12}},\ \bibinfo {pages}
			{1139} (\bibinfo {year} {2016})}\BibitemShut {NoStop}%
		\bibitem [{\citenamefont {Zhang}\ \emph {et~al.}(2024)\citenamefont {Zhang},
			\citenamefont {Zhang}, \citenamefont {Xu}, \citenamefont {Hu}, \citenamefont
			{Bao},\ and\ \citenamefont {Shen}}]{zhang_realizing_2024}%
		\BibitemOpen
		\bibfield  {author} {\bibinfo {author} {\bibfnamefont {Z.}~\bibnamefont
				{Zhang}}, \bibinfo {author} {\bibfnamefont {F.}~\bibnamefont {Zhang}},
			\bibinfo {author} {\bibfnamefont {Z.}~\bibnamefont {Xu}}, \bibinfo {author}
			{\bibfnamefont {Y.}~\bibnamefont {Hu}}, \bibinfo {author} {\bibfnamefont
				{H.}~\bibnamefont {Bao}},\ and\ \bibinfo {author} {\bibfnamefont
				{H.}~\bibnamefont {Shen}},\ }\href
		{https://doi.org/10.1103/PhysRevLett.133.133601} {\bibfield  {journal}
			{\bibinfo  {journal} {Phys. Rev. Lett.}\ }\textbf {\bibinfo {volume} {133}},\
			\bibinfo {pages} {133601} (\bibinfo {year} {2024})}\BibitemShut {NoStop}%
		\bibitem [{\citenamefont {Emary}\ and\ \citenamefont
			{Bishop}(2002)}]{emary_bogoliubov_2002}%
		\BibitemOpen
		\bibfield  {author} {\bibinfo {author} {\bibfnamefont {C.}~\bibnamefont
				{Emary}}\ and\ \bibinfo {author} {\bibfnamefont {R.~F.}\ \bibnamefont
				{Bishop}},\ }\href {https://doi.org/10.1063/1.1490406} {\bibfield  {journal}
			{\bibinfo  {journal} {J. Math. Phys.}\ }\textbf {\bibinfo {volume} {43}},\
			\bibinfo {pages} {3916} (\bibinfo {year} {2002})}\BibitemShut {NoStop}%
	\end{thebibliography}
	%apsrev4-2.bst 2019-01-14 (MD) hand-edited version of apsrev4-1.bst
	%Control: key (0)
	%Control: author (72) initials jnrlst
	%Control: editor formatted (1) identically to author
	%Control: production of article title (-1) disabled
	%Control: page (0) single
	%Control: year (1) truncated
	%Control: production of eprint (0) enabled
	%
	
\end{document}